\documentclass[floats,prd,aps,showpacs,amssymb,nofootinbib]{revtex4}
\usepackage{amssymb}
\usepackage{graphics}
\usepackage{amsmath}
\usepackage{amsfonts}
\usepackage{bm}
\usepackage[]{latexsym}

\newcommand{\be}{\begin{equation}}\newcommand{\ee}{\end{equation}}
\newcommand{\bea}{\begin{eqnarray}}\newcommand{\eea}{\end{eqnarray}}
\newcommand{\brr}{\begin{array}}\newcommand{\err}{\end{array}}
\newcommand{\bit}{\begin{itemize}}\newcommand{\eit}{\end{itemize}}
\newcommand{\ben}{\begin{enumerate}}\newcommand{\een}{\end{enumerate}}

\newcommand{\ba}{\begin{array}}
\newcommand{\ea}{\end{array}}

\def\lf{\left}

\def\ran{\rangle}

\def\ri{\right}
\def\al{\alpha}

\def\si{\sigma}
\def\Om{\Omega}

\def\1{{_{1}}}\def\2{{_{2}}}
\def\bk{{\bf {k}}}\def\bx{{\bf {x}}}

\def\noHe0{:\;\!\!\;\!\!:H_e(0):\;\!\!\;\!\!:}
\def\noHm0{:\;\!\!\;\!\!:H_\mu(0):\;\!\!\;\!\!:}

\def\lf{\left}

\def\ran{\rangle}

\def\ri{\right}

\def\al{\alpha}

\def\si{\sigma}
\def\Om{\Omega}

\def\1{{_{1}}}\def\2{{_{2}}}

\def\bogub{{\rho}^{\bf{k}}_{12}}
\def\bogvb{{\lambda}^{\bf{k}}_{12}}

\def\bogocoeffAlpha{{\cal A}_{(\Om,\Om'),\,\vec{k}}^{(\si,\si')}}

\def\bogocoeffBeta{{\cal B}_{(\Om,\Om'),\,\vec{k}}^{(\si,\si')}}

\def\bogocoeffBetater{{\cal B}_{(\Om,\Om''),\,\vec{k}}^{(\si,\si'')}}

\def\bogocoeffAlphater{{\cal A}_{(\Om,\Om''),\,\vec{k}}^{(\si,\si'')}}

\def\bogocoeffAlphaquintst{{\cal A}_{(\Om',\Om''),\,\vec{k}'}^{(\si',\si'')\,*}}
\def\bogocoeffBetaquintst{{\cal B}_{(\Om',\Om''),\,\vec{k}'}^{(\si',\si'')\,*}}
\def\Betasiomst{{\cal B}_{(\Om,\Om''),\,\vec{k}}^{(\si,\si')\,*}}
\def\Betasiom{{\cal B}_{(\Om',\Om''),\,\vec{k}'}^{(\si,\si')}}
\def\bogocoeffalphatilde{\widetilde{\alpha}_{\kappa\kappa'}^{\,(\sigma,\sigma')\,*}}
\def\bogocoeffbetatilde{\widetilde{\beta}_{\kappa\tilde\kappa'}^{\,(\sigma,\sigma')}}
\def\bogocoeffAlphamensigstarkapdue{{\cal A}_{(\Om,\Om'),\,\vec{k}}^{(\si,\si')}}
\def\bogocoeffAlphamensigstarkapduestarbi{{\cal A}_{(\Om',\Om),\,\vec{k}}^{(\si',\si)\,*}}
\def\bogocoeffBetamensigstarkapdue{{\cal B}_{(\Om,\Om'),\,\vec{k}}^{(\si,\si')}}
\def\bogocoeffBetamensigstarkapduestarb{{\cal B}_{(\Om',\Om),\,\vec{k}}^{(\si',\si)}}
\def\densityBB{N_{\mathcal{BB}}}

\def\densityAA{N_{\mathcal{AA}}}
\def\densityAB{N_{\mathcal{AB}}}
\def\densityBA{N_{\mathcal{BA}}}
\def\uuu{U}

\begin{document}
\title{Non-thermal signature of the Unruh effect in field mixing}

\author{M Blasone\footnote{blasone@sa.infn.it}$^{\hspace{0.3mm}1,2}$, G Lambiase\footnote{lambiase@sa.infn.it}$^{\hspace{0.3mm}1,2}$ and G G Luciano\footnote{gluciano@sa.infn.it}$^{\hspace{0.3mm}1,2}$} \affiliation
{$^1$Dipartimento di Fisica, Universit\'a di Salerno, Via Giovanni Paolo II, 132 I-84084 Fisciano (SA), Italy.\\ $^2$INFN, Sezione di Napoli, Gruppo collegato di Salerno, Italy.}

\date{\today}
  \def\be{\begin{equation}}
\def\ee{\end{equation}}
\def\al{\alpha}
\def\bea{\begin{eqnarray}}
\def\eea{\end{eqnarray}}

\begin{abstract}

Mixing transformations for a uniformly accelerated observer (Rindler observer) are analyzed within the quantum field theory framework as a basis for investigating gravitational effects on flavor oscillations. In particular, the case of two charged boson fields with different masses is discussed. In spite of such a minimal setting, the standard Unruh radiation is found to loose its characteristic thermal interpretation due to the interplay between the Bogolubov transformation hiding in field mixing and the one arising from the Rindler spacetime structure. The modified spectrum detected by the Rindler observer is explicitly calculated in the limit of small mass difference.
\end{abstract}

 \vskip -1.0 truecm
\maketitle

\section{Introduction}
\setcounter{equation}{0}
Since Pontecorvo's revolutionary idea \cite{Pontecorvo}, the theoretical basis of flavor mixing has been widely investigated. Although years of effort have been devoted to providing evidence for flavor oscillations, intriguing questions still remain open. Among these, for instance, the origin of this phenomenon within the Standard Model and the non-trivial condensate structure exhibited by the vacuum for mixed fields are the most puzzling problems. The latter aspect, in particular, has burst into the spotlight after the unitary inequivalence between mass and flavor vacua in the quantum field theory framework (QFT) was highlighted \cite{BV95,bosonmix}.

Flavor mixing in QFT is notoriously a non-trivial issue \cite{GiuntiKimLam}, since it is related with the problem of inequivalent representations of the canonical commutation relations \cite{bvbook}. The origin of this result lies in the fact that mixing transformations, which act as pure rotations on massive particle states in quantum mechanics (QM), have a more complicated structure at level of field operators. Indeed, they include both rotations and  Bogolubov transformations \cite{Gargiulo}, thereby inducing a condensate of particle/antiparticle pairs into the flavor vacuum. This has been pointed out first for Dirac fermions \cite{BV95} and later for other fields \cite{bosonmix,Blasone:2002jv,neutral}, showing in both cases the limits of the quantum mechanical approach in the treatment of flavor mixing.

All of the previous work has been carried out only in Minkowski spacetime. The existing literature on mixing and flavor oscillations in curved background, indeed, deals with this issue by using several other approaches, e.g. the WKB approximation \cite{Stodo, WudkaJ2001, Visinelli}, plane-wave method \cite{Kim, Konno} or geometric treatments \cite{Cardall}, and in various metrics, such as Schwarzschild \cite{Burgard, Piriz, Ahluwalia:1997hc, Kim}, Kerr \cite{Konno, WudkaJ2001}, Kerr-Newman, Friedmann--Robertson--Walker \cite{Visinelli}, Hartle-Thorne \cite{Geralico} and Lense-Thirring \cite{LambiasePapini} metrics. Thus it arises the question how the above formalism gets modified in the presence of gravity.

In the present work, a first step along this direction is taken by analyzing the QFT of two mixed scalar fields in a uniformly accelerated frame (Rindler metric). Despite such a minimal setting, a rich mathematical framework arises due to the combination of the Bogolubov transformation associated with mixing and the one related to the Rindler spacetime structure \cite{Unruh, Hawking, Fulling, Mukhanov, Birrell, Takagi:1986kn,Iorio:2002rr}. The extension of these results to the fermionic case, and in particular to neutrino fields, could provide new insights into the controversial problem of flavor oscillations in curved spacetime.

Mixing transformations in a non-inertial frame may serve as a tool for analyzing a number of other current questions dealing with such a topic: the spin-down of a rotating star by neutrino emission \cite{Dvornikov:2009rk} and the disagreement between the inverse $\beta$-decay rates of accelerated protons in comoving and inertial frames \cite{Aluw}, for instance, are some of the most relevant problems appearing in this framework. The latter aspect, in particular, is discussed, providing a possible resolution for the above incompatibility.

The paper is structured as follows: in the next Section, as a basis for extending mixing transformations to the Rindler frame, we introduce in Minkowski spacetime the hyperbolic field--quantization, that is, the scheme which diagonalizes Lorentz boost generator. The results obtained within such a framework are compared with the more familiar ones in plane-wave basis, thereby showing their equivalence for an inertial observer. In Section III the Rindler-Fulling quantization and the related Unruh effect are reviewed. Mixing transformations for the Rindler observer are derived in Section IV; the modified spectrum of Unruh radiation is explicitly calculated in the limit of small mass difference, thus showing its non--thermal nature when mixed fields are involved. Conclusions are briefly discussed in Section V. The paper is completed with three Appendices.

Throughout all the work, the metric $\eta_{\mu\nu}={\rm diag}(+1,-1,-1,-1)$ and natural units $\hslash=c=1$ will be used. In addition, the following notation for 4-, 3- and 2-vectors will be adopted
\begin{equation}
\label{eqn:notation}
x=\{t,\textbf{x}\},\qquad \bx=\{x^1,\vec{x}\},\qquad \vec{x}=\{x^2,x^3\}.
\end{equation}

\section{Quantized scalar field in hyperbolic representation}
\label{Hyperbrepr}

Let us consider a free complex scalar field $\phi$ with mass $m$ in a four-dimensional Minkowski spacetime. In the standard plane-wave representation, the field expansion reads
\begin{equation}
\phi(x)=\int d^{3}{k}\, \Big\{a_{\textbf{k}}\, \uuu_{\textbf{k}}(x)+ {\bar a_\textbf{k}}^\dagger\, \uuu_\textbf{k}^{\hspace{0.3mm}*}(x) \Big\},
\label{eqn:expans0}
\end{equation}
where
\begin{equation}
U_\bk(x)={\big[2\omega_{\bk}{(2\pi)}^{3}\big]}^{-\frac{1}{2}}\, e^{i\left(\bk\cdot\bx-\omega_{\bk} t\right)}.
\label{eqn:modes0}
\end{equation}
As known, the field quanta created by the application of the ladder operators $a_{\bk}^\dagger$ ($\bar a_{\bk}^\dagger$) on the Minkowski vacuum $|0_M\rangle$ carry well defined momentum $\bk$ and frequency $\omega_{\bk}={\sqrt{m^2+|\textbf{k}|^2}}$ with respect to the Minkowski time $t$ (see Appendix \ref{Plane--wave}). We will refer to these quanta as Minkowski particles (antiparticles), in contrast to the Rindler quanta to be later defined.

In order to extend the field--quantization scheme to the Rindler framework, let us introduce the so-called hyperbolic representation, that is, the representation which diagonalizes Lorentz boost operator. To check this, we look at the expression of the Lorentz-group generators:
\begin{equation}
M^{(\alpha,\beta)}=\int d^{3}x\, \left(x^{\alpha}\, T^{(0,\beta)}-x^{\beta}\, T^{(0,\alpha)} \right).
\label{eqn:gener}
\end{equation}
The boost operator (for example along the $x^1$ axis) is the $(1,0)$ component of $M^{(\alpha,\beta)}$. Using the standard expression of the stress tensor $T_{\mu\nu}$ and replacing the field expansion Eq. \eqref{eqn:expans0}, we obtain \cite{Zuber}
\begin{equation}
M^{(1, 0)}=i\int\frac{d^{3}k}{2\omega_\bk}\,\Big(c_\bk^{\, \dagger}\, \omega_\bk\, \frac{\partial}{\partial k_1}c_\bk+\bar c_\bk^{\, \dagger}\, \omega_\bk\, \frac{\partial}{\partial k_1}\bar c_\bk\Big),
\label{eqn:boostgenerator}
\end{equation}
where $c_\bk\equiv\sqrt{2\omega_\bk} \, a_\bk$. The result in Eq. (\ref{eqn:boostgenerator}) shows that $M^{(1, 0)}$ has a non-diagonal structure in the plane-wave representation. With a straightforward calculation, however, it can be verified that such a task is carried out by the following operators \cite{Takagi:1986kn}:
\begin{equation}
\label{eqn:operat-d}
d_{\kappa}^{\,(\sigma)}=\int^{+\infty}_{-\infty}\!\!\!\! dk_1\, p_\Omega^{\,(\sigma)}(k_1)\, a_{\bk},\qquad \bar d_{\kappa}^{\,(\sigma)}=\int^{+\infty}_{-\infty}\!\!\!\! dk_1\, p_\Omega^{\,(\sigma)}(k_1)\, \bar a_{\bk},
\end{equation}
where the subscript $\kappa$ stands for $(\Omega, \vec{k})$, $\sigma=\pm 1$, $\Omega$ is a positive parameter and\hspace{0.3mm}\footnote{The physical meaning of $\sigma$ and $\Omega$ will be explained in the next Section, where the quantization procedure is analyzed in a uniformly accelerated frame.}
\begin{equation}
p_\Omega^{\,(\sigma)}(k_1)=\frac{1}{\sqrt{2\pi\omega_\bk}}\, {\bigg(\frac{\omega_{\bk}+k_1}{\omega_{\bk}-k_1}\bigg)}^{i\sigma\Omega/2}\, .
\label{eqn:p}
\end{equation}
In terms of these operators, indeed, the boost generator $M^{(1, 0)}$ takes the form
\begin{equation}
M^{(1, 0)}\,=\,\int d^3 \kappa \sum_{\sigma}\sigma\,\Omega\left(d_{\kappa}^{\,(\sigma)\dagger}\,d_{\kappa}^{\,(\sigma)}\,+\,\bar d_{\kappa}^{\,(\sigma)\dagger}\,\bar d_{\kappa}^{\,(\sigma)}\right),
\end{equation}
which is clearly diagonal.

For later use, it is worth noting that the functions $p_\Omega^{\,(\sigma)}$ in Eq. (\ref{eqn:p}) form a complete orthonormal set, i.e.
\begin{eqnarray}
\label{eqn:completeorthonorm-p}
&&\hspace{7mm}\sum_{\sigma,\hspace{0.4mm}\Omega}\, p_\Omega^{\,(\sigma)}(k_1)\, p_\Omega^{\,(\sigma)*}(k_1')=\delta(k_1-k_1'),\\[2mm]
\label{eqn:compl-p}
&&\int_{-\infty}^{+\infty}\!\!\!\! dk_1\,\, p_{\Omega}^{\,(\sigma)*}(k_1)\, p_{\Omega'}^{ (\sigma')}(k_1)=\delta_{\sigma\sigma'}\delta(\Omega-\Omega'),
\end{eqnarray}
where the following shorthand notation has been introduced:
\begin{equation}
\sum_{\sigma,\hspace{0.4mm}\Omega}\equiv\sum_{\sigma}\int_{0}^{+\infty}\hspace{-2.0mm}d\Omega\hspace{0.2mm}.
\label{eqn:simpnot}
\end{equation}

Since the operators $d_{\kappa}^{\,(\sigma)}$ ($\bar d_{\kappa}^{\,(\sigma)}$) in Eq. (\ref{eqn:operat-d}) are linear combinations of the Minkowski annihilators  $a_{\bk}$ ($\bar a_{\bk}$) alone, they also annihilate the Minkowski vacuum $|0_M\rangle$ in Eq. (\ref{eqn:Minkvac}):
\begin{equation}
d_{\kappa}^{\,(\sigma)}\,|0_M\rangle=\bar d_{\kappa}^{\,(\sigma)}\,|0_M\rangle=0\hspace{0.2mm},\qquad \forall \sigma,\, \kappa.
\label{eqn:annihil-d}
\end{equation}
In addition, by exploiting Eqs. (\ref{eqn:completeorthonorm-p}), (\ref{eqn:compl-p}) and the commutation relations of $a_{\bk}$ and $\bar a_{\bk}$ in Eq. (\ref{eqn:commutrelations}), it is immediate to verify that the transformations Eq. (\ref{eqn:operat-d}) are canonical, i.e.
\bea
\label{eqn:commut-d}
\left[d_{\kappa}^{\,(\sigma)}\,,\, d_{\kappa'}^{\,(\sigma')\dagger}\Big]=\Big[\bar d_{\kappa}^{\,(\sigma)}\,,\, \bar d_{\kappa'}^{\,(\sigma')\dagger}\right]=\delta_{\sigma\sigma'}\, \delta^3(\kappa-\kappa'),
\eea
with all other commutators vanishing. Therefore, Eqs. (\ref{eqn:annihil-d}) and (\ref{eqn:commut-d}) allow us to state that, from the viewpoint of an inertial observer, the hyperbolic and plane-wave quantizations are equivalent at level of ladder operators.

The hyperbolic wave functions associated with the operators $d_{\kappa}^{\,(\sigma)}$  can be now derived by inverting Eq. (\ref{eqn:operat-d}) with respect to $a_{\bk}$ and $\bar a_{\bk}$ and substituting the resulting expressions into the field expansion Eq. \eqref{eqn:expans0}. It follows that
\begin{equation}
\phi(x)=\sum_{\sigma,\hspace{0.4mm}\Omega}\,\int d^{2}k\,\Big\{d_{\kappa}^{\,(\sigma)}\;\widetilde{U}_{\kappa}^{\,(\sigma)}(x) + \bar d_{\kappa}^{\,(\sigma)\dagger}\;\widetilde{U}_{\kappa}^{\,(\sigma)*}(x)\Big\},
\label{eqn:expansionfieldutilde}
\end{equation}
where
\begin{equation}
\widetilde{U}_{\kappa}^{\,(\sigma)}(x)=\int_{-\infty}^{+\infty}\!\!\!\! dk_1\, p_\Omega^{\,(\sigma)*}(k_1)\,  U_\bk(x)\hspace{0.2mm}.
\label{eqn:Uwidetilde}
\end{equation}
The integral Eq. (\ref{eqn:Uwidetilde}) can be more directly solved by introducing the Rindler coordinates $(\eta,\xi)$, related to the Minkowski ones by the following expressions
\begin{equation}
\label{eqn:rindlercoordinates}
t=\xi\sinh\eta\,,\qquad x^1=\xi\cosh\eta\,,
\end{equation}
with $-\infty<\eta,\xi<\infty$ (note that $x^2$ and $x^3$ are common to both sets of coordinates). We have\footnote{The set of coordinates $(\eta,\xi,\vec{x})$ in Eq. (\ref{eqn:wideutilde}) is denoted by $x$, as well as the corresponding set of Minkowski coordinates $(t,x^1,\vec{x})$ in Eq. (\ref{eqn:modes0}). Therefore, according to our convention, the symbol $x$ refers to a spacetime point, rather than its representation in a particular coordinate system.} (see Ref.\cite{Proceeding})
\begin{equation}
\label{eqn:wideutilde}
\widetilde{U}_{\kappa}^{\,(\sigma)}(x)=\frac{\,e^{\sigma\pi\Omega/2}}{2 \sqrt{2}\,\pi^{2}}
\,K_{i\sigma\Omega}(\mu_k\xi)\,e^{i\left(\vec{k}\cdot\vec{x}-\sigma\Omega\eta\right)},
\end{equation}
where $K_{i\sigma\Omega}(\mu_k\xi)$ is the modified Bessel function of second kind and $\mu_{k}$ is the reduced frequency:
\be
\mu_{k}=\sqrt{m^2+|\vec{k}|^2},
\label{eqn:muk}
\ee
with $\vec{k}\equiv \{k^2,k^3\}$ as introduced in Eq. (\ref{eqn:notation}).

It is not difficult to show that the hyperbolic modes in Eq. (\ref{eqn:Uwidetilde}) form a complete orthonormal set with respect to the KG inner product Eq. (\ref{eqn:prodscal}). i.e.
\bea
\Big(\widetilde{U}_{\kappa'}^{\,(\sigma')}\, ,\widetilde{U}_{\kappa}^{\,(\sigma)}\Big)=-\Big(\widetilde{U}_{\kappa'}^{\,(\sigma')*}\, ,\widetilde{U}_{\kappa}^{\,(\sigma)*}\Big)=\delta_{\sigma\sigma'}\delta^3(\kappa-\kappa'), \qquad \Big(\widetilde{U}_{\kappa'}^{\,(\sigma')}\, ,\widetilde{U}_{\kappa}^{\,(\sigma)*}\Big)=0\hspace{0.2mm}.
\label{eqn:ortonormbis}
\eea

Before turning to discuss the quantization procedure for an accelerated observer, it should be emphasized that, although the plane-wave expansion Eq. (\ref{eqn:expans0}) applies to all the points of spacetime, the hyperbolic representation Eq. (\ref{eqn:expansionfieldutilde}) is valid only on the Rindler manifolds $x^1>|t|\hspace{0.4mm}\cup\hspace{0.4mm} x^1<-|t|$. By analytically continuing the solutions (\ref{eqn:wideutilde}) across $x_1=\pm\, t$, one obtains the correct global functions, i.e. the Gerlach's Minkowski Bessel modes (see Ref.\cite{Gerlach}). For our purpose, nevertheless, it is enough to consider the modes as above defined.

\section{Field--quantization in a uniformly accelerated frame: Unruh effect}
\label{Unruh}
The above--discussed hyperbolic representation provides a springboard for analyzing the Rindler-Fulling quantization in a uniformly accelerated frame \cite{Fulling}.  As a first step for such an extension, by exploiting the Rindler coordinates $(\eta, \xi, x^2, x^3)$ in Eq. (\ref{eqn:rindlercoordinates}), let us rewrite the line element $ds^2=\eta_{\mu\nu}dx{^\mu} dx{^\nu}$ in the form
\begin{equation}
ds^2={(dt)}^2-{(dx^1)}^2-\sum_{j=2}^{3}{(dx^j)}^2\underset{\tiny{Rindler\; coord.}}{\longrightarrow}ds^2=\xi^2d\eta^2-d\xi^2-\sum_{j=2}^{3}{(dx^j)}^2\,.
\label{eqn:lineelement}
\end{equation}
Since the metric does not depend on $\eta$, the vector $B=\frac{\partial }{\partial\eta}$ is a timelike Killing vector. By exploiting Eq. (\ref{eqn:rindlercoordinates}), one can verify that $B$ coincides with the boost Killing vector along the $x^1$ axis.

The physical relevance of the Rindler coordinates can be readily explained by considering the following world line
\begin{equation}
\xi(\tau)=\mathrm{const}\equiv a^{-1}\,,\quad x^2(\tau)=\mathrm{const}\,,\quad x^3(\tau)=\mathrm{const},
\label{eqn:lineadiuni}
\end{equation}
where $\tau$ is the proper time measured along the line. By inserting Eq. (\ref{eqn:lineadiuni}) into the metric  Eq. (\ref{eqn:lineelement}), we find that
\begin{equation}
\eta(\tau)=a\tau.
\label{eqn:rindlertime}
\end{equation}
Therefore, the proper time $\tau$ of an observer along the line (\ref{eqn:lineadiuni}) is the same as the Rindler time $\eta$, up to the scale factor $a$. We will refer to such an observer as Rindler observer.

Equation \eqref{eqn:rindlertime} plays a striking role; according to the above discussion on the Killing vector $\frac{\partial }{\partial\eta}$, indeed, it shows that the time evolution for the Rindler observer is properly an infinite succession of infinitesimal Minkowski boost transformations. This is the reason why in the first Section we deeply insisted on the hyperbolic representation as opposed to the more familiar plane-wave field expansion.

\begin{figure}[t]
\resizebox{9cm}{!}{\includegraphics{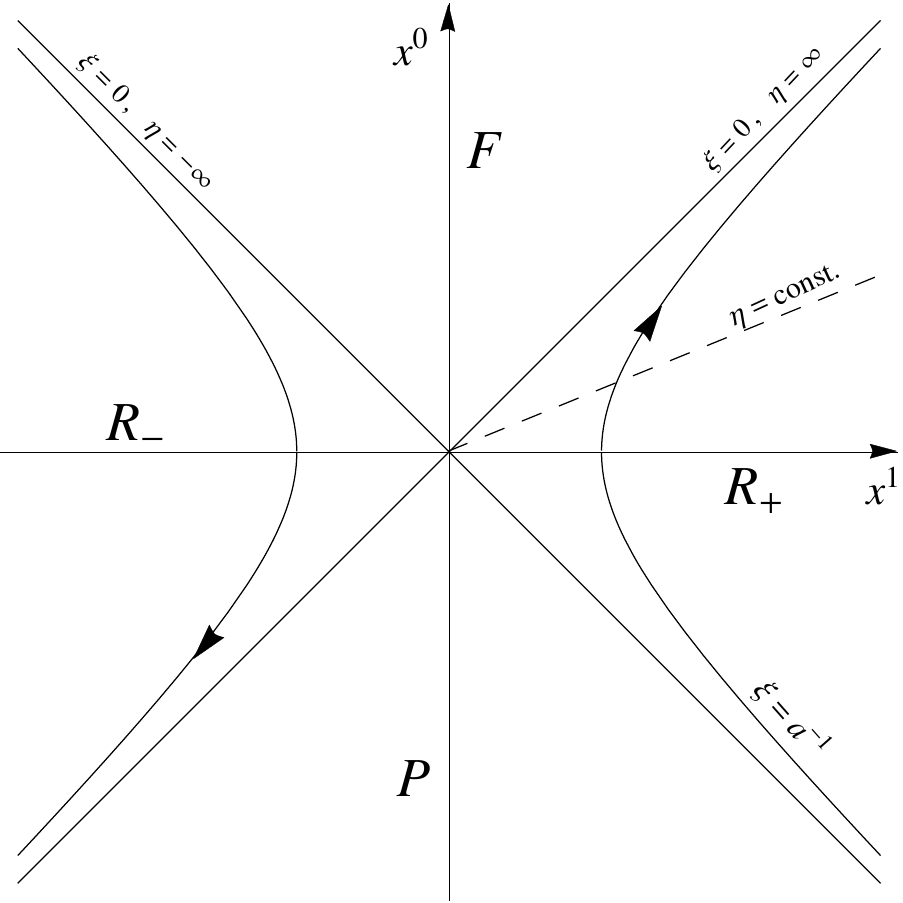}}
\caption{\small{The proper coordinate system of a uniformly accelerated observer in the Minkowski spacetime.The hyperbola represents the world line of an observer with proper acceleration $a$.
}}
\label{figure:Rindler}
\end{figure}

In the Minkowski coordinates $(t, x^1, x^2, x^3)$, the world line Eq. (\ref{eqn:lineadiuni}) takes the form
\begin{equation}
t(\tau)=a^{-1}\sinh a\tau,\quad x^1(\tau)=a^{-1}\cosh a\tau,\quad x^2(\tau)=\mathrm{const},\quad x^3(\tau)=\mathrm{const}.
\label{eqn:lineauniversomink}
\end{equation}
Equation \eqref{eqn:lineauniversomink} describes an hyperbola in the $(t,x^1)$ plane with asymptotes $t=\pm\, x^1$ (Fig.\ref{figure:Rindler}). It is not difficult to see that it represents the world line of a uniformly accelerated observer with proper acceleration $|a|$ \cite{Mukhanov}; for $a>0$, in particular, the observer is confined within the right wedge $R_+=\{x|x^1>|t|\}$, for $a<0$, conversely, his motion occurs in the left wedge $R_-=\{x|x^1<-|t|\}$. On this basis, the physical difference between the Minkowski and Rindler metrics can be pointed out; as shown in Fig.\ref{figure:Rindler}, a uniformly accelerated observer in $R_+$ is causally separated from one in $R_-$. Indeed, he cannot receive (send) any signal from the future (past) wedge $t > |x^1|$ $(t < -|x^1|) $. Therefore, the null hyperplane $t=|x^1|$ ($t=-|x^1|$) appears to him as a future (past) event horizon. The above considerations, however, do not apply to the Minkowski (inertial) observer, whose signals, sent or received, can reach every point of spacetime.

We are now ready to describe the field--quantization procedure from the viewpoint of a uniformly accelerated observer. In Appendix \ref{Klein--Gordon equation in Rindler coordinates} the solutions of the Klein--Gordon equation in Rindler coordinates are explicitly derived:
\begin{equation}
u_\kappa^{\,(\sigma)}(x)=\theta(\sigma\xi)\!\ {\left[2\Omega{(2\pi)^{2}}\right]}^{-\frac{1}{2}}\!\ h_\kappa^{\,(\sigma)}(\xi)\!\ e^{i\left(\vec{k}\cdot\vec{x}-\sigma\Omega\eta\right)},
\label{eqn:rindlermodesbis}
\end{equation}
where $\sigma=\pm 1$ refers to the right/left wedges $R_\pm$, $\Omega$ is the frequency with respect to the Rindler time $\eta$ and $h_\kappa^{\,(\sigma)}$ is the modified Bessel function of second kind, up to a normalization factor (see Eq. (\ref{eqn:besselmodified})). The Heaviside step function $\theta(\sigma\xi)$ has been inserted in Eq. (\ref{eqn:rindlermodesbis}) in order to restrict the Rindler modes  $u_\kappa^{\,(\sigma)}$ to only one of the two causally separated wedges $R_\pm$.

Exploiting the completeness and orthonormality properties of the set $\{u_\kappa^{\,(\sigma)},\,u_{\kappa}^{\,(\sigma)\,*}\}$, we can expand the field in the Rindler framework as follows
\begin{equation}
\phi(x)=\sum_{\sigma,\hspace{0.4mm}\Omega}\,\int d^{2}k\,\Big\{{b^{\,(\sigma)}_\kappa}\, u_\kappa^{\,(\sigma)}(x)+{\bar b^{\,(\sigma)\dagger}_\kappa}\, u_\kappa^{\,(\sigma)*}(x) \Big\},
\label{eqn:espanrind}
\end{equation}
where $\kappa\equiv(\Omega, \vec{k})$ as already defined. The ladder operators ${b^{\,(\sigma)}_\kappa}$ and ${\bar b^{\,(\sigma)}_\kappa}$ are assumed to obey the canonical commutation relations:
\begin{equation}
\left[b_\kappa^{\,(\sigma)}, b_{\kappa'}^{ ( \sigma')\dagger}\Big]=\Big[{\bar b_\kappa}^{\,(\sigma)},\, {\bar b_{\kappa'}}^{\,(\sigma')\dagger}\right]=\delta_{\sigma\sigma'}\,\delta^3(\kappa-\kappa')\hspace{0.2mm},
\label{eqn:commutcanon2}
\end{equation}
with all other commutators vanishing. They can be interpreted as annihilation operators of Rindler--Fulling particles and antiparticles, respectively.  The Rindler--Fulling vacuum, denoted with $|0_R\rangle$, is accordingly defined by
\begin{equation}
 b_\kappa^{\,(\sigma)}|0_R\rangle=\bar b_\kappa^{\,(\sigma)}|0_R\rangle=0,\qquad \forall\sigma, \kappa.
\label{eqn:vuotodirindler}
\end{equation}

In order to figure out the connection between the Minkowski and Rindler quantizations, let us now compare the two alternative field--expansions on a spacelike hypersurface $\Sigma$ lying in the Rindler manifolds $R_\pm$ (for instance, we may consider an hyperplane of constant $\eta$). Due to the equivalence of plane--wave and hyperbolic representations within the Minkowski framework, we could equally consider the relations Eqs. (\ref{eqn:expans0}) and (\ref{eqn:expansionfieldutilde}) for the inertial observer. To simplify the calculations, we opt for the latter. Therefore, by equating  Eqs. (\ref{eqn:expansionfieldutilde}) and (\ref{eqn:espanrind}) on the hypersurface $\Sigma$ and forming the KG inner product of both sides with the Rindler mode $u_\kappa^{\,(\sigma)}$, we have
\begin{equation}
b_{\kappa}^{\,(\sigma)}=\sqrt{\big(1+N_R\big(\Omega\big)\big)}\!\ d_{\kappa}^{\, (\sigma)}+\sqrt{N_R\big(\Omega\big)}\,\bar d_{\tilde\kappa}^{\!\ (-\sigma)\dagger}\,,
\label{eqn:newformbogotransform}
\end{equation}
where $\tilde\kappa\equiv(\Omega,-\vec{k})$ and
\begin{equation}
N_R(\Omega)=\frac{1}{e^{2\pi\Omega}-1}
\label{eqn:distrib}
\end{equation}
is the Bose--Einstein distribution function. We will refer to Eq. (\ref{eqn:newformbogotransform}) as  \lq\lq thermal\rq\rq\  Bogolubov transformation.

To be complete, let us observe that, if we used the plane--wave expansion Eq. (\ref{eqn:expans0}) instead of the hyperbolic scheme, the transformation Eq. (\ref{eqn:newformbogotransform}) would take the far less \lq\lq manageable\rq\rq\hspace{0.1mm} form
\begin{equation}
{b^{\,(\sigma)}_{\kappa}}=\int d^3{\bk'}\, \left\{a_{\bk'}\,{\alpha^{\,(\sigma)*}_{\kappa\bk'}}+{\bar a_{\bk'}}^\dagger\,{\beta^{\,(\sigma)*}_{\kappa\bk'}}\right\},
\label{eqn:bogotransform}
\end{equation}
with $a_{\bk'}$ and $\bar a_{\bk'}$ introduced in Eq. (\ref{eqn:expans0}) and:
\bea
\label{eqn:bogocoeff}
&&{\alpha^{\,(\sigma)}_{\kappa\bk'}}=\frac{e^{\pi\Omega/2}\, |\Gamma(i\Omega)|}{2\pi}{\bigg(\frac{\Omega}{\omega_{\bk'}}\bigg)}^{1/2}{\bigg(\frac{\omega_{\bk'}+k_1'}{\omega_{\bk'}-k_1'}\bigg)}^{-i\sigma\Omega/2}\,\delta^2(\vec{k}-\vec{k}'),
\\[4mm]
&&{\beta^{\,(\sigma)}_{\kappa\bk'}}=\frac{e^{-\pi\Omega/2}\, |\Gamma(i\Omega)|}{2\pi}{\bigg(\frac{\Omega}{\omega_{\bk'}}\bigg)}^{1/2}{\bigg(\frac{\omega_{\bk'}+k_1'}{\omega_{\bk'}-k_1'}\bigg)}^{-i\sigma\Omega/2}\,\delta^2(\vec{k}+\vec{k}').
\eea

The distribution of Rindler particles in the Minkowski vacuum can be now readily calculated by exploiting Eqs. (\ref{eqn:annihil-d}) and (\ref{eqn:newformbogotransform}). It follows that
\begin{equation}
\langle0_M|b_{\kappa}^{\,(\sigma)\dagger}\!\ b_{\kappa'}^{ (\sigma)}|0_M\rangle= N_R(\Omega)\!\ \delta^3{(\kappa-\kappa')},
\label{eqn:aspectval}
\end{equation}
which is clearly non--vanishing. Moreover, by noting from Eq. (\ref{eqn:rindlertime}) that the proper energy of the particles seen by a Rindler observer with acceleration $a$ is $a\Omega$, it can be verified that the condensate Eq. (\ref{eqn:aspectval}) has a thermal spectrum, with temperature $T$ given by
\begin{equation}
T=\frac{a}{2\pi}\,.
\label{eqn:T}
\end{equation}
One can easily recognize in Eqs. (\ref{eqn:aspectval}), (\ref{eqn:T}) the well--known Unruh effect, which states that, from the viewpoint of the Rindler observer, inertial vacuum appears as a thermal bath with temperature proportional to the magnitude of his acceleration.

\section{Flavor mixing transformations for an accelerated observer}
\label{flmixrind}
In the last two decades mixing transformations in QFT have been widely investigated first for fermions \cite{BV95} and then for bosons \cite{bosonmix}, showing in both cases the presence of non--trivial vacuum structure for the flavor fields. However, these studies have been carried out only for an inertial observer in the plane--wave and hyperbolic representations \cite{Castiglioncello}. Thus it arises the question how the above structure appears in general frame and, in particular, from the viewpoint of the Rindler observer. To this end, starting from the review in Appendix \ref{Mixing}, let us consider mixing relations in a simplified two--flavor model:
\begin{eqnarray}
\label{Pontecorvoa}
\phi_{A}(x)&=&\phi_{1}(x)\, \cos\theta  + \phi_{2}(x)\, \sin\theta ,\\ [2mm]
\label{Pontecorvob}
\phi_{B}(x) &=& -  \phi_{1}(x)\, \sin\theta   +  \phi_{2}(x)\, \cos\theta,
\end{eqnarray}
where $\phi_i$, $i=1,2$ are two free complex scalar fields with masses $m_i$, $\phi_\chi$, $\chi=A,B$ are the mixed fields and $\theta$ is the mixing angle. Following the same approach as in Section \ref{Unruh}, in the Rindler framework we adopt the following free fields--like expansions for mixed fields\hspace{0.3mm}\footnote{For simplicity, the time dependence of the flavor operators will be omitted when there is no ambiguity.}
\bea
\label{eqn:expancampmixrind}
\phi_{A}(x)&=& \sum_{\sigma,\hspace{0.4mm}\Omega}\,\int d^{2}k\,\Big\{b_{\kappa,A}^{\,(\sigma)}\;{u}_{\kappa,1}^{\,(\sigma)}(x)+\bar b_{\kappa,A}^{\,(\sigma)\dagger}\;{u}_{\kappa,1}^{\,(\sigma)*}(x)\Big\} ,\label{eqn:sub expancampmixrindler a}
\\[1mm]
\phi_{B}(x)&=&\sum_{\sigma,\hspace{0.4mm}\Omega}\,\int d^{2}k\,\Big\{ b_{\kappa,B}^{\,(\sigma)}\;{u}_{\kappa,2}^{\,(\sigma)}(x)+\bar b_{\kappa,B}^{\,(\sigma)\dagger}\;{u}_{\kappa,2}^{\,(\sigma)*}(x)\Big\},\label{eqn:sub expancampmixrindler b}
\eea
where $b_{\kappa,\chi}^{\,(\sigma)}$, $\bar b_{\kappa,\chi}^{\,(\sigma)}$ and their respective adjoints ($\chi=A,B$) are the flavor operators for the Rindler observer.

As pointed out in Appendix \ref{Mixing}, mixing relations in QFT inherently hide a Bogolubov transformation relating the ladder operators for flavor fields with the corresponding ones for definite mass fields (see Eq. (\ref{eqn:flavor-a})). Therefore, by virtue of Eq. (\ref{eqn:newformbogotransform}), we expect that $b_{\kappa,\chi}^{\,(\sigma)}$ and $\bar b_{\kappa,\chi}^{\,(\sigma)}$ in Eqs. (\ref{eqn:sub expancampmixrindler a}), (\ref{eqn:sub expancampmixrindler b}) are related to the \lq\lq mass\rq\rq\hspace{0.2mm}  operators for an inertial observer by the combination of two Bogolubov transformations, the one arising from the Rindler spacetime structure, the other associated with flavor mixing.

In order to analyze such an interplay, let us equate the (hyperbolic) Minkowski representation of $\phi_A$ in Eq. (\ref{eqn:sub_expancampmix_a})  with the corresponding Rindler expansion Eq. (\ref{eqn:sub expancampmixrindler a}):
\begin{equation}
\sum_{\sigma,\hspace{0.4mm}\Omega}\,\int d^{2}k\,\Big\{b_{\kappa,A}^{\,(\sigma)}\;{u}_{\kappa,1}^{\,(\sigma)}(x)+\bar b_{\kappa,A}^{\,(\sigma)\dagger}\;{u}_{\kappa,1}^{\,(\sigma)*}(x)\Big\}= \sum_{\sigma,\hspace{0.4mm}\Omega}\,\int d^{2} {k}\,\Big\{d_{\kappa,A}^{\,(\sigma)}\;\widetilde{U}_{\kappa,1}^{\,(\sigma)}(x)+\bar d_{\kappa,A}^{\,(\sigma)\dagger}\;\widetilde{U}_{\kappa,1}^{\,(\sigma)*}(x)\Big\},
\end{equation}
where we emphasize that $d_{\kappa, A}^{\,(\sigma)}$ and $\bar d_{\kappa, A}^{\,(\sigma)}$ are the flavor operators for an inertial observer within the hyperbolic scheme.
By multiplying both sides for the Rindler mode $u_{\kappa,1}^{\,(\sigma)}$ and using the orthonormality condition Eq. (\ref{eqn:rindlernorm1}), we have
\begin{equation}
b_{\kappa,A}^{\,(\sigma)}=\sum_{\sigma',\hspace{0.4mm}\Omega'}\,\int d^{2}k'\,\Big\{d_{\kappa', A}^{\,(\sigma')}\;\bogocoeffalphatilde+\,\bar d_{\tilde\kappa',A}^{\, (-\sigma')\dagger}\;\bogocoeffbetatilde\,\Big\},
\label{eqn:relationbd}
\end{equation}
where
\begin{equation}
\label{eqn:hypercoeff}
\widetilde{\alpha}_{\kappa\kappa'}^{(\sigma,\sigma')}=\Big(u_{\kappa,1}^{\,(\sigma)},\,\widetilde{U}_{\kappa',1}^{\,(\sigma')}\Big),
\qquad\bogocoeffbetatilde=\Big(\widetilde{U}_{\tilde\kappa',1}^{\, (-\sigma')*},\, u_{\kappa,1}^{\,(\sigma)}\Big).
\end{equation}
The Bogolubov coefficients $\widetilde{\alpha}_{\kappa\kappa'}^{(\sigma,\sigma')}$ and $\bogocoeffbetatilde$ in Eq. (\ref{eqn:hypercoeff}) are clearly independent of the mixing angle $\theta$: therefore, by noting that the transformation Eq. (\ref{eqn:relationbd}) must reduce to Eq. (\ref{eqn:newformbogotransform}) for $\theta\rightarrow0$ (since $b_{\kappa,A}^{\,(\sigma)}\big|_{{}_{\theta=0}}\rightarrow b_{\kappa,1}^{\,(\sigma)}$,\, $d_{\kappa,A}^{\,(\sigma)}\big|_{{}_{\theta=0}}\rightarrow d_{\kappa,1}^{\,(\sigma)}$ and $d_{\tilde\kappa,A}^{\, (-\sigma)}\big|_{{}_{\theta=0}}\rightarrow d_{\tilde\kappa,1}^{\, (-\sigma)}$\,), we readily obtain  (up to an irrelevant global phase factor\hspace{0.3mm}\footnote{In our treatment the phase factor turns out to be irrelevant since we want to calculate the expectation value $\langle0_M|b_{\kappa, A}^{\,(\sigma)\dagger}\, b_{\kappa',A}^{ (\sigma')}|0_M\rangle$.})\hspace{0.2mm}
\be
b_{\kappa,A}^{\,(\sigma)}=\sqrt{(1+N_R(\Omega))}\; d_{\kappa,A}^{\,(\sigma)}+\sqrt{N_R(\Omega)}\; \bar d_{\tilde\kappa,A}^{\, (-\sigma)\dagger}\, .
\label{eqn:newformbogotransform2}
\ee
Similarly, for $\bar b_{\kappa,A}^{\,(\sigma)}$ we have
\be
\bar b_{\kappa,A}^{\,(\sigma)}=\sqrt{(1+N_R(\Omega))}\; \bar d_{\kappa,A}^{\,(\sigma)}+\sqrt{N_R(\Omega)}\;  d_{\tilde\kappa,A}^{\, (-\sigma)\dagger}\, .
\label{eqn:newformbogotransform3}
\ee

The corresponding relation between $b_{\kappa,B}^{\,(\sigma)}$ and $d_{\kappa,B}^{\,(\sigma)}$ can be derived by equating the expansions of $\phi_B$ in Eqs.  (\ref{eqn:sub expancampmixrindler b}) and (\ref{eqn:sub expancampmix b})  and forming the inner product of both sides with $u_{\kappa,2}^{\,(\sigma)}$. A straightforward calculation leads to
\be
b_{\kappa,B}^{\,(\sigma)}=\sqrt{(1+N_R(\Omega))}\; d_{\kappa,B}^{\,(\sigma)}+\sqrt{N_R(\Omega)}\; \bar d_{\tilde\kappa,B}^{\, (-\sigma)\dagger}\,.
\label{eqn:newformbogotransform2b}
\ee
Similarly, for $\bar b_{\kappa,B}^{\,(\sigma)}$ we have
\be
\bar b_{\kappa,B}^{\,(\sigma)}=\sqrt{(1+N_R(\Omega))}\; \bar d_{\kappa,B}^{\,(\sigma)}+\sqrt{N_R(\Omega)}\;  d_{\tilde\kappa,B}^{\, (-\sigma)\dagger}\,.
\label{eqn:newformbogotransform3b}
\ee

Using the transformations Eq. (\ref{eqn:newformbogotransform2}) and (\ref{eqn:newformbogotransform2b}), we can now calculate the Rindler spectrum of mixed--particles in the inertial vacuum. To this purpose, however, the explicit expressions of $d_{\kappa,\chi}^{\,(\sigma)}$ and $\bar d_{\kappa,\chi}^{\,(\sigma)}$ in Eqs. (\ref{eqn:daoperator})-(\ref{eqn:dbbaroperator}) are required:
\begin{eqnarray}
\label{eqn:da1}
d_{\kappa,A}^{\,(\sigma)}&=&\cos\theta\, d_{\kappa,1}^{\,(\sigma)}
\,+\,\sin\theta\sum_{\sigma',\hspace{0.4mm}\Omega'}\,\Big(d_{(\Omega',\vec{k}),2}^{\,(\sigma')}\;\bogocoeffAlpha \,+\,\bar d_{(\Omega',-\vec{k}),2}^{\,(\sigma')\dagger}\;\bogocoeffBeta\,\Big),\\[2mm]
\label{eqn:dabar1}
\bar d_{\kappa,A}^{\,(\sigma)}&=& \cos\theta\, \bar d_{\kappa,1}^{\,(\sigma)}
\,+\,\sin\theta\sum_{\sigma',\hspace{0.4mm}\Omega'}\,\Big(\bar d_{(\Omega',\vec{k}),2}^{\,(\sigma')}\;\bogocoeffAlpha\, +\,d_{(\Omega',-\vec{k}),2}^{\,(\sigma')\dagger}\;\bogocoeffBeta\,\Big),\\[2mm]
\label{eqn:db1}
d_{\kappa,B}^{\,(\sigma)}&=&\cos\theta\, d_{\kappa,2}^{\,(\sigma)}
\,-\,\sin\theta\sum_{\sigma',\hspace{0.4mm}\Omega'}\,\Big(d_{(\Omega',\vec{k}),1}^{\,(\sigma')}\;\bogocoeffAlphamensigstarkapduestarbi \, -\,\bar d_{(\Omega',-\vec{k}),1}^{\,(\sigma')\,\dagger}\;\bogocoeffBetamensigstarkapduestarb\,\Big),\\[2mm]
\label{eqn:dbbar1}
\bar d_{\kappa,B}^{\,(\sigma)}&=&\cos\theta\, \bar d_{\kappa,2}^{\,(\sigma)}
-\,\sin\theta\sum_{\sigma',\hspace{0.4mm}\Omega'}\,\Big(\bar d_{(\Omega',\vec{k}),1}^{\,(\sigma')}\;\bogocoeffAlphamensigstarkapduestarbi \, -\, d_{(\Omega',-\vec{k}),1}^{\,(\sigma')\,\dagger}\;\bogocoeffBetamensigstarkapduestarb\,\Big),
\end{eqnarray}
where the mixing Bogolubov coefficients $\bogocoeffAlpha$ and $\bogocoeffBeta$ are given by Eqs. (\ref{eqn:coefficientuno}) and (\ref{eqn:coefficientdue})
\begin{eqnarray}
&& \bogocoeffAlpha =\int^{+\infty}_{-\infty}\frac{dk_1}{4\pi}\lf(\frac{1}{\omega_{\bk,1}}+\frac{1}{\omega_{\bk,2}}\ri){\lf(\frac{\omega_{\bk,1}+k_1}{\omega_{\bk,1}-k_1}\ri)}^{i\sigma\Omega/2} {\lf(\frac{\omega_{\bk,2}+k_1}{\omega_{\bk,2}-k_1}\ri)}^{-i\sigma'\Omega'/2}
e^{i(\omega_{\bk,1}-\,\omega_{\bk,2})t}\,,
\label{eqn:coefficientunobis}
\\ [4mm]
&& \bogocoeffBeta =\int^{+\infty}_{-\infty}\frac{dk_1}{4\pi}\lf(\frac{1}{\omega_{\bk,2}}
-\frac{1}{\omega_{\bk,1}}\ri){\lf(\frac{\omega_{\bk,1}+
k_1}{\omega_{\bk,1}-k_1}\ri)}^{i\sigma\Omega/2} {\lf(\frac{\omega_{\bk,2}+k_1}{\omega_{\bk,2}-k_1}\ri)}^{-i\sigma'\Omega'/2}
e^{i(\omega_{\bk,1}+\,\omega_{\bk,2})t}\,.
\label{eqn:coefficientduebis}
\end{eqnarray}
As discussed in Appendix \ref{Mixing}, the analytical resolution of these integrals is non--trivial. Further developments in the calculation of the modified Unruh distribution, however, can be obtained for  $t=\eta=0$; indeed, in this case, by exploiting the relations Eqs. (\ref{eqn:coefficientunobistzeroappendix}), (\ref{eqn:coefficientduebistzeroappendix}), the Bogolubov transformation Eq.  (\ref{eqn:newformbogotransform2}) can be recast in the form
\begin{eqnarray}
\label{eqn:bogtransfbis}
b_{\kappa,A}^{\,(\sigma)}&=&\sqrt{1+N_R(\Omega)}\,\,
\Big[\cos\theta\, d_{\kappa,1}^{\,(\sigma)}
\,+\,\sin\theta\hspace{-0.6mm}\sum_{\hspace{0.6mm}\sigma',\hspace{0.1mm}\Omega'}\Big(d_{(\Omega',\vec{k}),2}^{\,(\sigma')}\,\;\bogocoeffAlpha \,+\,\bar d_{(\Omega',-\vec{k}),2}^{\,(\sigma')\dagger}\,\;\bogocoeffBeta\,\Big)\Big]\\[2mm]
&+&\,\sqrt{N_R(\Omega)}\,\,\Big[\cos\theta\,\bar d_{\tilde\kappa,1}^{\,(-\sigma)\dagger}
\,+\,\sin\theta\hspace{-0.6mm}\sum_{\hspace{0.6mm}\sigma',\hspace{0.1mm}\Omega'}\Big(\bar d_{(\Omega',-\vec{k}),2}^{(-\sigma')\dagger}\,\;\bogocoeffAlphamensigstarkapdue\,+\,d_{(\Omega',\vec{k}),2}^{(-\sigma')}\,\;\bogocoeffBetamensigstarkapdue\,\Big)\Big],
\end{eqnarray}
where, to simplify the notation, the subscript $t=0$ has been omitted. Therefore, the spectrum of mixed--particles detected by the Rindler observer in the inertial vacuum takes the form
\bea
\label{eqn:divergenceexpectvalue}
\hspace{-15.5mm}{\cal N}_{R}(\theta)\Big|_0&\equiv& \langle0_M|\,b_{\kappa,\chi}^{\,(\sigma)\dagger}\,
b_{\kappa',\chi}^{\,(\sigma)}\,|0_M\rangle\\[2mm]\nonumber
\hspace{-5mm}&=&\,N_R(\Omega)\,\cos^2\theta\,\delta^3{(\kappa-\kappa')}\,+\,\sin^2\theta\Bigg[\sqrt{N_R(\Omega)}\,\sqrt{N_R(\Omega')}\,\densityAA\,+\,\sqrt{1+N_R{(\Omega)}}\,\sqrt{1+N_R(\Omega')}\,\densityBB\\[2mm]\nonumber
&+&\,\sqrt{1+N_R(\Omega)}\,\sqrt{N_R(\Omega')}\,\densityBA\,+\,\sqrt{N_R{(\Omega)}}\,\sqrt{1+N_R(\Omega')}\,\densityAB\Bigg]\,\delta^2(\vec{k}-\vec{k}'),\qquad \chi=A,B,
\eea
where $N_R(\Omega)$ is the standard Unruh condensate Eq. (\ref{eqn:distrib}) and the following notation has been introduced
\begin{eqnarray}
&& \densityAA\equiv\sum_{\hspace{0.6mm}\sigma',\hspace{0.1mm}\Omega''}{{\cal A}_{(\Om,\Om''),\,\vec{k}}^{(\si,\si')\,*}}\;\,{{\cal A}_{(\Om',\Om''),\,\vec{k}'}^{(\si,\si')}},\qquad N_{\mathcal{BB}}\equiv\sum_{\hspace{0.6mm}\sigma',\hspace{0.1mm}\Omega''}\Betasiomst\;\,\Betasiom,\\[3mm]
&&\densityAB\equiv\sum_{\hspace{0.6mm}\sigma',\hspace{0.1mm}\Omega''}{{\cal A}_{(\Om,\Om''),\,\vec{k}}^{(\si,\si')\,*}}\;\,{{\cal B}_{(\Om',\Om''),\,\vec{k}'}^{(\si,-\si')}},\qquad \densityBA\equiv\sum_{\hspace{0.6mm}\sigma',\hspace{0.1mm}\Omega''}{{\cal B}_{(\Om,\Om''),\,\vec{k}}^{(\si,\si')\,*}}\;\,{{\cal A}_{(\Om',\Om''),\,\vec{k}'}^{(\si,-\si')}}\hspace{0.2mm}.
\end{eqnarray}
From Eq. (\ref{eqn:divergenceexpectvalue}), by using the relations Eqs. (\ref{eqn:primprop}), (\ref{eqn:secprop}) and defining:
\begin{eqnarray}
\label{eqn:F}
F(\Omega,\Omega')&\equiv&\sqrt{N_R(\Omega)\,N_R(\Omega')}\,+\,\sqrt{\left(1+N_R(\Omega)\right)\left(1+N_R(\Omega')\right)}\,,\\[3mm]
G(\Omega,\Omega')&\equiv&\sqrt{1+N_R(\Omega)}\,\sqrt{N_R(\Omega')}\,+\,\sqrt{N_R(\Omega)}\,\sqrt{1+N_R(\Omega')}\,,
\label{eqn:G}
\end{eqnarray}
it follows that
\begin{eqnarray}
\label{eqn:finalresult}
{\cal N}_{R}(\theta)\Big|_0\,=\,N_R(\Omega)\,\delta^3(\kappa-\kappa')\,+\,\sin^2\theta\Big[F(\Omega,\Omega')\,\densityBB\,+\,G(\Omega,\Omega')\,\densityAB\Big]\,\delta^2(\vec{k}-\vec{k}'),
\end{eqnarray}
which coincides with the result obtained in Ref.\cite{Castiglioncello} by directly expressing the flavor $d$--operators in Eqs. (\ref{eqn:sub_expancampmix_a}), (\ref{eqn:sub expancampmix b}) in terms of the corresponding ones in plane--wave representation Eqs. ({\ref{eqn:sub expansion bis a}}), ({\ref{eqn:sub expansion bis b}}).

Therefore, due to the interplay between mixing and thermal Bogolubov transformations, the radiation detected by the Rindler observer  gets significantly modified, resulting in the sum of the conventional Unruh density plus non--diagonal corrections arising from flavor mixing. Let us separately evaluate these additional terms; to this end, by exploiting Eqs. (\ref{eqn:completeorthonorm-p}), (\ref{eqn:compl-p}), the densities $N_{\mathcal{BB}}$ and $N_{\mathcal{AB}}$ in Eq. (\ref{eqn:finalresult}) can be rewritten as
\begin{eqnarray}
\label{eqn:nbb}
N_{\mathcal{BB}}&=&-\frac{1}{2}\,\delta\,(\Omega-\Omega')\,+\,J^{(-\Omega,\Omega')}\Big(\frac{1}{\omega_{\bk,2}}\Big)\,+\, K^{(-\Omega,\Omega')} \Big(\frac{\omega_{\bk,2}}{\omega_{\bk,1}^{\hspace{0.3mm}2}}\Big),\\[2mm]
\label{eqn:nab}
\hspace{12mm}N_{\mathcal{AB}}&=&J^{(-\Omega,-\Omega')}\Big(\frac{1}{\omega_{\bk,2}}\Big)\,-\, K^{(-\Omega,-\Omega')} \Big(\frac{\omega_{\bk,2}}{\omega_{\bk,1}^{\hspace{0.3mm}2}}\Big),
\end{eqnarray}
where
\begin{eqnarray}
\label{eqn:F}
J^{(-\Omega,\Omega')}\Big(\frac{1}{\omega_{\bk,2}}\Big)&\equiv&\int^{+\infty}_{-\infty}\frac{dk_1}{8\pi\omega_{\bk,2}}\,{\lf(\frac{\omega_{\bk,1}+k_1}{\omega_{\bk,1}-k_1}\ri)}^{-i\sigma\Omega/2} {\lf(\frac{\omega_{\bk,1}+k_1}{\omega_{\bk,1}-k_1}\ri)}^{i\sigma\Omega'/2},\\[4mm]
\label{eqn:G}
K^{(-\Omega,\Omega')} \Big(\frac{\omega_{\bk,2}}{\omega_{\bk,1}^{\hspace{0.3mm}2}}\Big)&\equiv&\int^{+\infty}_{-\infty}\frac{dk_1}{8\pi}\,\frac{\omega_{\bk,2}}{\omega_{\bk,1}^{\hspace{0.3mm}2}}\,{\lf(\frac{\omega_{\bk,1}+k_1}{\omega_{\bk,1}-k_1}\ri)}^{-i\sigma\Omega/2} {\lf(\frac{\omega_{\bk,1}+k_1}{\omega_{\bk,1}-k_1}\ri)}^{i\sigma\Omega'/2}.
\end{eqnarray}
We focus on $N_{\mathcal{BB}}$; the calculation of $N_{\mathcal{AB}}$ is obviously similar. In the reasonable limit of small mass difference $\frac{\Delta m^2}{m_i^2}\ll 1$, $i=1,2$, by using the approximations Eqs. (\ref{eqn:approx1}), (\ref{eqn:approx2}), the integrals in Eqs. (\ref{eqn:F}) and (\ref{eqn:G}) take the form
\begin{eqnarray*}
\label{eqn:Fappr}
&&\hspace{-6mm}J^{(-\Omega,\Omega')}\hspace{0.2mm}=\hspace{0.2mm}
\frac{1}{4}\,\delta\,(\Omega-\Omega')-\int^{+\infty}_{-\infty}\frac{dk_1}{8\pi}\left(\frac{1}{2}\,\frac{\Delta m^2}{\omega_{\bk,1}^{\hspace{0.3mm}3}}\hspace{0.3mm}-\hspace{0.3mm}\frac{3}{8}\,\frac{{(\Delta m^2)}^2}{\omega_{\bk,1}^{\hspace{0.3mm}5}}\right){\lf(\frac{\omega_{\bk,1}+k_1}{\omega_{\bk,1}-k_1}\ri)}^{-i\sigma\Omega/2} {\lf(\frac{\omega_{\bk,1}+k_1}{\omega_{\bk,1}-k_1}\ri)}^{i\sigma\Omega'/2}+
\mathcal{O}\left({\Big(\frac{\Delta m^2}{\mu_{k,1}^{\hspace{0.3mm}2}}\Big)}^3\right),\\[4mm]
\label{eqn:Gappr}
&&\hspace{-6mm}K^{(-\Omega,\Omega')}\hspace{0.2mm}=\hspace{0.2mm}\frac{1}{4}\,\delta\,(\Omega-\Omega')+\int^{+\infty}_{-\infty}\frac{dk_1}{8\pi}\left(\frac{1}{2}\,\frac{\Delta m^2}{\omega_{\bk,1}^{\hspace{0.3mm}3}}\hspace{0.2mm}-\hspace{0.2mm}\frac{1}{8}\,\frac{{(\Delta m^2)}^2}{\omega_{\bk,1}^{\hspace{0.3mm}5}}\right){\lf(\frac{\omega_{\bk,1}+k_1}{\omega_{\bk,1}-k_1}\ri)}^{-i\sigma\Omega/2} {\lf(\frac{\omega_{\bk,1}+k_1}{\omega_{\bk,1}-k_1}\ri)}^{i\sigma\Omega'/2}+\hspace{0.2mm}
\mathcal{O}\left({\Big(\frac{\Delta m^2}{\mu_{k,1}^{\hspace{0.3mm}2}}\Big)}^3\right),
\end{eqnarray*}
where $\mu_{k,1}$ is defined in Eq. (\ref{eqn:muk}). Therefore, for $N_{\mathcal{BB}}$ we obtain
\begin{eqnarray}
\label{eqn:Nbbaprox}
N_{\mathcal{BB}}&=&\frac{1}{4}\,\int^{+\infty}_{-\infty}\frac{dk_1}{8\pi}\frac{{(\Delta m^2)}^2}{\omega_{\bk,1}^{\hspace{0.3mm}5}}\,{\lf(\frac{\omega_{\bk,1}+k_1}{\omega_{\bk,1}-k_1}\ri)}^{-i\sigma\Omega/2} {\lf(\frac{\omega_{\bk,1}+k_1}{\omega_{\bk,1}-k_1}\ri)}^{i\sigma\Omega'/2}\,+\,
\mathcal{O}\left({\Big(\frac{\Delta m^2}{\mu_{k,1}^{\hspace{0.3mm}2}}\Big)}^3\right)\\[4mm]
\nonumber
&=&\frac{1}{192}\;\frac{{(\Delta m^2)}^2}{\mu_{k,1}^{\hspace{0.3mm}4}}\;\frac{\sigma\,(\Omega'-\Omega)}{\sinh\left[\frac{\pi}{2}\,\sigma\,(\Omega'-\Omega)\right]}\left[4+{(\Omega'-\Omega)}^2\right]\,+\,
\mathcal{O}\left({\Big(\frac{\Delta m^2}{\mu_{k,1}^{\hspace{0.3mm}2}}\Big)}^3\right),
\end{eqnarray}
where in the last step we used the formula (see Ref.\cite{Abram}):
\begin{equation}
\int_{-\infty}^{+\infty}\frac{dy}{\cosh^4y}\, \cos\left[\sigma\,(\Omega'-\Omega)\,y\right]\,=\,\frac{1}{6}\,\frac{\pi\,\sigma\,(\Omega'-\Omega)}{\sinh\left[\frac{\pi}{2}\,\sigma\,(\Omega'-\Omega)\right]}\left[4+{(\Omega'-\Omega)}^2\right].
\end{equation}
Similarly, for $N_{\mathcal{AB}}$ it can be verified that
\begin{eqnarray}
\label{eqn:Nabaprox}
N_{\mathcal{AB}}&=&\int^{+\infty}_{-\infty}\frac{dk_1}{8\pi}\left(-\,\frac{\Delta m^2}{\omega_{\bk,1}^{\hspace{0.3mm}3}}\,+\,\frac{1}{2}\frac{{(\Delta m^2)}^2}{\omega_{\bk,1}^{\hspace{0.3mm}5}}\right)\,{\lf(\frac{\omega_{\bk,1}+k_1}{\omega_{\bk,1}-k_1}\ri)}^{-i\sigma\Omega/2} {\lf(\frac{\omega_{\bk,1}+k_1}{\omega_{\bk,1}-k_1}\ri)}^{-i\sigma\Omega'/2}\,+\,\mathcal{O}\left({\Big(\frac{\Delta m^2}{\mu_{k,1}^{\hspace{0.3mm}2}}\Big)}^3\right)\\[4mm]
\nonumber
&=&-\frac{1}{8}\,\frac{\Delta m^2}{\mu_{k,1}^{\hspace{0.3mm}2}}\,\frac{\sigma\,(\Omega'+\Omega)}{\sinh\left[\frac{\pi}{2}\,\sigma\,(\Omega'+\Omega)\right]}\,+\,\frac{1}{96}\;\frac{{(\Delta m^2)}^2}{\mu_{k,1}^{\hspace{0.3mm}4}}\;\frac{\sigma\,(\Omega'+\Omega)}{\sinh\left[\frac{\pi}{2}\,\sigma\,(\Omega'+\Omega)\right]}\left[4+{(\Omega'+\Omega)}^2\right]\,+\,\mathcal{O}\left({\Big(\frac{\Delta m^2}{\mu_{k,1}^{\hspace{0.3mm}2}}\Big)}^3\right),
\end{eqnarray}
where the formula Eq. (\ref{eqn:formula}) has been used. In definitive, by inserting Eqs. (\ref{eqn:Nbbaprox}), (\ref{eqn:Nabaprox}) in Eq. (\ref{eqn:finalresult}), it follows that
\begin{eqnarray}
\label{eqn:spectrumapprox}
&&\hspace{-7mm}{\cal N}_{R}(\theta)\Big|_0\,=\,N_R(\Omega)\,\delta^3(\kappa-\kappa')-\,\sin^2\theta\,\Bigg\{\frac{\Delta m^2}{8\,\mu_{k,1}^{\hspace{0.3mm}2}}\;G(\Omega,\Omega')\,\frac{\sigma\,(\Omega'+\Omega)}{\sinh\left[\frac{\pi}{2}\,\sigma\,(\Omega'+\Omega)\right]}\\[3mm]
\nonumber
&&\hspace{6.8mm}+\,\,\frac{{(\Delta m^2)}^2}{96\,\mu_{k,1}^{\hspace{0.3mm}4}}\,\bigg[\frac{F(\Omega,\Omega')}{2}\,\frac{\sigma\,(\Omega'-\Omega)}{\sinh\left[\frac{\pi}{2}\,\sigma\,(\Omega'-\Omega)\right]}\left[4+{(\Omega'-\Omega)}^2\right]+\,G(\Omega,\Omega')\,\frac{\sigma\,(\Omega'+\Omega)}{\sinh\left[\frac{\pi}{2}\,\sigma\,(\Omega'+\Omega)\right]}\left[4+{(\Omega'+\Omega)}^2\right]\bigg]\\[3mm]
\nonumber
&&\hspace{6.8mm}+\,\,\mathcal{O}\left({\Big(\frac{\Delta m^2}{\mu_{k,1}^{\hspace{0.3mm}2}}\Big)}^3\right)\Bigg\}\,\delta\,{(\vec{k}-\vec{k})},
\end{eqnarray}
with $F(\Omega,\Omega')$ and $G(\Omega,\Omega')$ defined in Eqs. (\ref{eqn:F}) and (\ref{eqn:G}), respectively.

Three comments are in order here. First, for $\theta\rightarrow 0$ Eq. (\ref{eqn:spectrumapprox}) correctly reduces to the Bose--Einstein distribution in Eq. (\ref{eqn:aspectval}), as one would expect in absence of mixing. Similar considerations hold for $m_1\rightarrow m_2$ and in the relativistic limit $|\vec{k}|^{2}\gg m_{1}^2+m_{2}^2$, since the parameter $\frac{\Delta m^2}{\mu^2_{k,i}}$ approaches zero.

Second, the total number of mixed particles with frequency $\Omega$ and $2$--momentum $\vec{k}$ can be obtained by integrating Eq. (\ref{eqn:spectrumapprox}) over $\kappa'$. It is not difficult to verify that the higher the frequency, the more relevant the contribution of the mixing corrections becomes.

Third, we emphasize that although the characteristic Unruh distribution Eq. (\ref{eqn:distrib}) does not  depend on the mass of the field detected by the Rindler observer, the modified spectrum Eq. (\ref{eqn:spectrumapprox}) turns out to be proportional to the squared mass difference of the two mixed fields. Therefore, flavor mixing breaks the mass--scale invariance of Unruh effect; as a consequence, vacuum--radiation looses its original thermal interpretation. Of course, the extension of such a result to the neutrino case can be potentially exploited to fix new constraints on the squared mass differences of these fields.
 
Borrowing Hawking's idea about black hole evaporation, here we sketch an heuristic interpretation of the above modification: in absence of mixing, as shown, inertial vacuum appears as a condensate of Rindler particle/antiparticle pairs all of the same type. Normally, these pairs exist for an extremely short time before annihilating. Nevertheless, just outside the event horizon, it is possible for an antiparticle to fall back into the Rindler inaccessible region before the annihilation occurs, in which case its partner can escape, observed as Unruh radiation. In other terms, for free fields we can state that the thermal bath detected by the accelerated observer in the inertial vacuum originates from the corresponding flux of one--type antiparticles crossing the horizon.

\begin{figure}[t]
\resizebox{9cm}{!}{\includegraphics{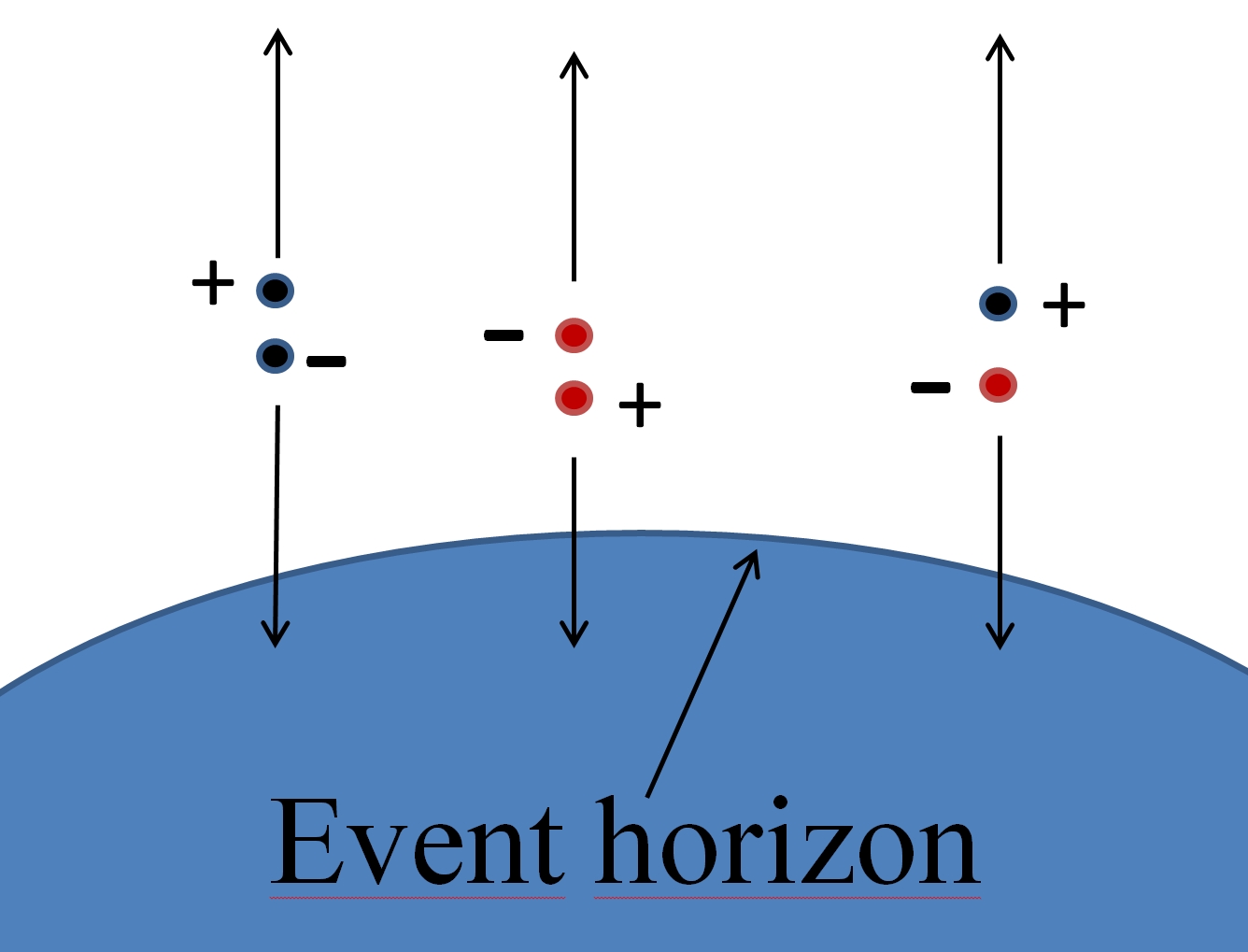}}
\caption{\small{Pictorial interpretation of the modified spectrum of mixed particles in  $|0_M\rangle$ for a Rindler observer. Different online--colours of dots correspond to different particle/antiparticle flavors. Contrary to the case of absence of mixing, vacuum is populated by particle--antiparticle pairs, both of the same (blue--blue, red--red) and different (blue--red) type.} }
\label{figure:unruhmix}
\end{figure}

The above considerations, however, get modified when mixed fields are involved. In this case, indeed, we stress that vacuum is populated by particle/antiparticle pairs both of the same and different flavors (see Refs.\cite{BV95,bosonmix} for details). Therefore, Unruh radiation can be generated by both types of antiparticles crossing the horizon. In other words, if a $B$--flavor particle escapes, it could correspond to a $B$-flavor antiparticle fallen back into the horizon, as well as to an $A$-flavor antiparticle (see Fig.\ref{figure:unruhmix}). Such an ambiguity, of course, will increase the entropy of system, thus modifying the spectrum of Unruh radiation.

\section{Discussion}
In this paper the topic of flavor mixing from the viewpoint of a uniformly accelerated observer has been analyzed within the quantum field theory framework. In particular, the case of two charged scalar fields with different masses has been discussed. Due to the combination of the two Bogolubov transformations involved -- the one hiding in flavor mixing, the other associated with the Rindler spacetime structure -- the spectrum of Unruh radiation is found to be significantly modified, resulting in the sum of the standard Bose--Einstein distribution plus non--trivial corrections arising from mixing (see Eq. (\ref{eqn:finalresult})). The explicit calculation of these additional terms has been performed in the limit of small mass difference, thereby showing that Unruh radiation looses its characteristic thermality when mixed fields are involved (Eq.  (\ref{eqn:spectrumapprox})).

In spite of its minimal setting, we stress that the quantization formalism developed in this paper provides a convenient starting point for analyzing flavor oscillations in the QFT on curved background. Our approach, indeed, once extended to the fermionic case, and in particular to neutrino fields, may give new insights into the existing formalisms dealing with this topic \cite{Stodo, Visinelli, Konno, Cardall}. Moreover, the problems of the phase shift between two neutrino mass eigenstates \cite{Visinelli, Burgard, Koranga} and neutrino spin oscillations in gravitational fields \cite{Dvornikov:2006ji} could be investigated.

Analyzing field mixing in an accelerated frame may serve as a basis for studying a number of other theoretical problems appearing in such a framework. For instance, it has been recently shown \cite{Aluw} that the inverse $\beta$--decay rates of accelerated protons in the inertial and comoving frames disagree in the context of neutrino flavor mixing.  As pointed out in Ref.\cite{Aluw}, such an incompatibility can be seen as the price of maintaining the Kubo--Martin--Schwinger (KMS) condition of thermal state for the accelerated neutrino vacuum. Nevertheless, according to what we have shown, this argument is flawed, since it requires this state to be thermal even in presence of mixed (interacting) fields.

Actually, the above result should not be surprising; flavor mixing, indeed, is not the only context in which the violation of the KMS condition in Rindler frame occurs. A similar situation is discussed in Ref.\cite{Hossain}, where the two--point function along Rindler trajectory is found to loose its thermal interpretation within the loop quantization method, due to  non--Lorentz invariance of polymer correction terms.

Along this line, a further interesting question to be potentially investigated is Lorentz invariance violation in the context of mixed neutrinos (see Ref.\cite{Di Mauro} for a more detailed treatment). Up to now, indeed, flavor mixing in the quantum field theory framework has been analyzed only within the usual plane--wave representation. Exploiting the hyperbolic scheme discussed in Section \ref{Hyperbrepr}, one could explore more efficiently such an issue.

It is our purpose to investigate this and other aspects in forthcoming papers. More work, indeed, is inevitably required along these lines.

\appendix
\section{Plane--wave representation in Minkowski spacetime}
\label{Plane--wave}
In this Appendix the standard plane--wave quantization of a free scalar field in Minkowski spacetime is briefly reviewed. As a starting point for such an analysis, by using the Minkowski coordinates $\{t,\bx\}\equiv\{t,x^1,x^2,x^3\}$, we expand the field in the familiar form
\begin{equation}
\phi(x)=\int d^{3}{k}\, \Big\{a_{\textbf{k}}\, \uuu_{\textbf{k}}(x)+ {\bar a_\textbf{k}}^\dagger\, \uuu_\textbf{k}^{\hspace{0.3mm}*}(x) \Big\},
\label{eqn:expans}
\end{equation}
where
\begin{equation}
U_\bk(x)={\big[2\omega_{\bk}{(2\pi)}^{3}\big]}^{-\frac{1}{2}}\, e^{i\left(\bk\cdot\bx-\omega_{\bk} t\right)}
\label{eqn:modes}
\end{equation}
are the plane--waves of frequency
\begin{equation}
\omega_{\bk}={\sqrt{m^2+|\textbf{k}|^2}}\,.
\label{eqn:freq&mom}
\end{equation}
The modes $U_\bk$ in Eq. (\ref{eqn:modes}) are solutions of the Klein--Gordon equation:
\begin{equation}
\bigg\{{\bigg(\frac{\partial}{\partial t}\bigg)}^2-\sum_{j=1}^{3}{\bigg(\frac{\partial}{\partial x^j}\bigg)}^2+m^2 \bigg\}\,\phi(x)=0\hspace{0.2mm}.
\label{eqn:esplicitKleinGordon}
\end{equation}
They are normalized with respect to the Klein--Gordon (KG) inner product:
\begin{equation}
\Big(\phi_1,\phi_2\Big)=i\int d^{3}{x}\, \Big[\phi_2^*(x)\stackrel{\leftrightarrow}{\partial_{t}}\phi_1(x)\Big],
 \label{eqn:prodscal}
\end{equation}
where the integration is assumed to be performed on a hypersurface of constant $t$. Indeed, we have
\bea
\label{eqn:minkowskinorm}
\Big(U_\bk,U_{\bk'}\Big)=-\Big(U^*_\bk,U^*_{\bk'}\Big)=\delta^3(\bk-\bk'), \quad\,\, \Big(U_\bk,U^*_{\bk'}\Big)=0\hspace{0.2mm}.
\eea
The operators $a_\bk$ and $\bar a_\bk$ in the field expansion Eq. \eqref{eqn:expans} are assumed to satisfy the canonical commutation relations:
\bea
\label{eqn:commutrelations}
&\Big[a_\bk, a^\dagger_{\bk'}\Big]=\Big[\bar a_\bk,\bar a^\dagger_{\bk'}\Big]=\delta^{3}(\bk-\bk'),
\eea
with all other commutators vanishing. As well known, they can be interpreted as annihilation operators of Minkowski particles and antiparticles, respectively. The Minkowski vacuum $|0_M\rangle$ is accordingly defined by
\begin{equation}
a_\bk|0_M\rangle=\bar a_\bk|0_M\rangle=0,\qquad \forall \bk\hspace{0.2mm}.
\label{eqn:Minkvac}
\end{equation}
In terms of $a_{\bk}$, $\bar a_{\bk}$ and their respective adjoints, it is easy to show that the (normal ordered) Hamiltonian and momentum operator have a diagonal structure \cite{Greiner}:
\begin{equation}
\label{eqn:Hamiltonian}
H=\int d^{3}{k}\,\omega_k\,\Big(a^\dagger_\bk a_\bk+\bar a^\dagger_\bk\bar a_\bk\Big),\qquad\quad \textbf P=\int d^{3}{k}\,\bk\,\Big(a^\dagger_\bk a_\bk+\bar a^\dagger_\bk\bar a_\bk\Big)\hspace{0.2mm}.
\end{equation}
Therefore, in the plane--wave representation, field quanta are characterized by well--defined momentum and energy with respect to the time $t$.

\section{Klein--Gordon equation for the Rindler observer}
\label{Klein--Gordon equation in Rindler coordinates}
In order to extend the field--quantization formalism to an accelerated frame, it is useful to rewrite the Klein--Gordon equation (\ref{eqn:esplicitKleinGordon}) in terms of the Rindler coordinates $\{\eta,\xi,x^2,x^3\}$ given in Eq. (\ref{eqn:rindlercoordinates}):
\begin{equation}
\bigg\{{\bigg(\frac{\partial}{\partial t}\bigg)}^2-\sum_{j=1}^{3}{\bigg(\frac{\partial}{\partial x^j}\bigg)}^2+m^2 \bigg\}\hspace{0.1mm}\,\phi(x)=0\underset{Rindler\;coord.}{\longrightarrow}
\bigg\{\frac{1}{\xi^2}\frac{\partial^2}{\partial\eta^2}-\frac{\partial^2}{\partial\xi^2}-\frac{1}{\xi}\frac{\partial}{\partial\xi}-\sum_{j=2}^{3}{\bigg(\frac{\partial}{\partial x^j}\bigg)}^2+m^2 \bigg\}\hspace{0.1mm}\,\phi(x)=0\,.
\label{eqn:esplicitKleinGordonrindler}
\end{equation}
Solutions of positive frequency $\Omega$ with respect to the Rindler time $\eta$ can be written in the form (see Ref.\cite{Takagi:1986kn})
\begin{equation}
u_\kappa^{\,(\sigma)}(x)=\theta(\sigma\xi)\!\ {\left[2\Omega{(2\pi)^{2}}\right]}^{-\frac{1}{2}}\!\ h_\kappa^{\,(\sigma)}(\xi)\!\ e^{i\left(\vec{k}\cdot\vec{x}-\sigma\Omega\eta\right)}\,,
\label{eqn:rindlermodes}
\end{equation}
where $\sigma=+$ refers to the right wedge  $R_+=\{x|x^1>|t|\}$, while $\sigma=-$ to the left wedge $R_-=\{x|x^1<-|t|\}$. We will refer to these functions as Rindler modes. Since the Rindler regions $R_{\pm}$ are causally separated from each other, the Heaviside step function $\theta(\sigma\xi)$ has been inserted in Eq. (\ref{eqn:rindlermodes}). The time dependence
\begin{equation}
u_\kappa^{\,(\sigma)}\propto e^{-i\sigma\Omega\eta}\,
\end{equation}
reflects the fact that the boost Killing vector $B=\frac{\partial}{\partial\eta}$ is future oriented in $R_+$, while it is past oriented in $R_-$.

The explicit expression of $h_\kappa^{\,(\sigma)}$ can be obtained by substituting Eq. (\ref{eqn:rindlermodes}) into Eq. (\ref{eqn:esplicitKleinGordonrindler}). This leads to
\begin{equation}
\bigg\{\frac{d^2}{d\xi^2}+\frac{1}{\xi}\frac{d}{d\xi}+\frac{\Omega^2}{\xi^2}-\mu_k^2 \bigg\}\,h_\kappa^{\,(\sigma)}(\xi)=0\hspace{0.2mm},
\label{eqn:sostitu}
\end{equation}
which is solved by the modified Bessel functions of second kind. In detail, by requiring that these functions are delta-normalized, one gets
\begin{equation}
h_\kappa^{\,(\sigma)}={\left(2/\pi\right)}^{\frac{1}{2}}A_\kappa^{\,(\sigma)}{\left(\alpha\mu_{k}/2\right)}^{i\Omega}\!\ \Gamma(i\Omega)^{-1}K_{i\Omega}(\mu_{k}\xi),
\label{eqn:besselmodified}
\end{equation}
with
\begin{equation}
\label{eqn:fattoredifase}
\quad A_{\kappa}^{\,(\sigma)}=\left\{ \begin{array}{l}
R^{\, *}_{\kappa}{\left(\alpha\mu_k/2\right)}^{-i\Omega}\, \Gamma(i\Omega)/|\Gamma(i\Omega)|, \qquad \mathrm{for}\, \,  \sigma=+\, ,
\\[2mm]
R_{\kappa}\,{\left(\alpha\mu_k/2\right)}^{i\Omega}\, \Gamma(-i\Omega)/|\Gamma(i\Omega)|, \quad\;\;\;\hspace{-0.3mm}\mathrm{for}\, \,  \sigma=-\, , \\
\end{array}\right.
\end{equation}
where $R_{\kappa}={\left[{\left(\alpha\mu_k/2\right)}^{-i\Omega}\, \Gamma(i\Omega)/|\Gamma(i\Omega)| \right]}^2$, $\alpha$ is an arbitrary postive constant of dimension of lenght and $\Gamma(i\Omega)$ is the Euler Gamma function.

By exploiting Eqs. (\ref{eqn:rindlermodes}), (\ref{eqn:besselmodified}), it is possible to verify that the Rindler modes above defined form a complete orthonormal set with respect to the KG inner product in Rindler coordinates:
\begin{equation}
\left(\phi_1,\phi_2\right)=i \int^{+\infty}_{-\infty}\frac{d\xi}{|\xi|}\int d^{2}{x}\hspace{0.8mm} \phi_2^*\!\ \overset{\leftrightarrow}\partial_ \eta \!\ \phi_1,
\label{eqn:proscalrindlercoord}
\end{equation}
where we have implicitly assumed that the integration is performed on a hypersurface of constant $\eta$. we have
\begin{equation}
\left(u_\kappa^{\,(\sigma)},\,u_{\kappa'}^{\,(\sigma')}\right)=-\left(u_\kappa^{\,(\sigma)*},\, u_{\kappa'}^{\,(\sigma)*}\right)=\delta_{\sigma\sigma'}\!\ \delta^3(\kappa-\kappa')\hspace{0.2mm},\qquad
\left(u_\kappa^{\,(\sigma)},\,u_{\kappa'}^{\,(\sigma)*}\right)=0\hspace{0.2mm}.
\label{eqn:rindlernorm1}
\end{equation}

\section{Boson field mixing in Minkowski spacetime: plane-wave and hyperbolic modes}
\label{Mixing}
In this Appendix flavor mixing transformations within the QFT framework are reviewed  \cite{bosonmix}. The background we consider for our discussion is the usual Minkowski spacetime. The results and definitions reported here provide the basis for extending such a formalism to the Rindler frame.

Let us start by considering mixing relations for scalar fields in a simplified model with only two flavors:
\bea \label{Ponteca}
\phi_{A}(x)&=& \phi_{1}(x)\, \cos\theta  + \phi_{2}(x)\, \sin\theta\hspace{0.2mm} ,\\ [2mm]
\label{Pontecb}
\phi_{B}(x) &=&-  \phi_{1}(x)\, \sin\theta   +  \phi_{2}(x)\, \cos\theta\hspace{0.2mm} ,
\eea
where $\phi_{\chi}$, $\chi=A,B$ denote the mixed fields and $\theta$ is the mixing angle. $\phi_i(x)$, $i=1,2$, are two free complex scalar fields with masses $m_i$, whose expansions in the plane-wave basis are given by Eq. (\ref{eqn:expans})
\begin{eqnarray}
\label{eqn:exp1}
&\phi_1(x)&=\int d^{3}{k}\, \Big\{a_{\textbf{k},1}\, \uuu_{\textbf{k},1}(x)+ {\bar a_{\textbf{k},1}}^\dagger\, {\uuu^*_{\textbf{k},1}}(x) \Big\},\\[1mm]
&\phi_2(x)&=\int d^{3}{k}\, \Big\{a_{\textbf{k},2}\, \uuu_{\textbf{k},2}(x)+ {\bar a_{\textbf{k},2}}^\dagger\, {\uuu^*_{\textbf{k},2}}(x) \Big\}.
\label{eqn:exp2}
\end{eqnarray}
The respective conjugate momenta are $\pi_i(x)=\partial_0\phi_i^\dagger(x)$
and their commutation relations are
\begin{equation}
\qquad\big[\phi_i(x),\,\pi_j(x')\big]_{t=t'}=\big[\phi^\dagger_i(x),\,\pi^\dagger_j(x')\big]_{t=t'}=i\,\delta^3(x-x')\,\delta_{ij}\,,\qquad i,j=1,2\,,
\label{eqn:relcommcan}
\end{equation}
with the other equal-time commutators vanishing.

Since we are dealing with two free fields, Minkowski vacuum is generalized as
\begin{equation}
\label{eqn:newvacuum}
|0_M\ran \equiv |0_M\ran_1\otimes|0_M\ran_2\,,
\end{equation}
 where $|0_M\ran_i$ is the vacuum for the field with mass $m_i$ (see the definition in Eq. (\ref{eqn:Minkvac}) or, equivalently, Eq. (\ref{eqn:annihil-d})).

The completeness of the two sets of plane-waves $\big\{{U}_{\bk,i}\, ,{U}_{\bk,i}^{*}\big\}$, $i=1,2$  allows us to take the following expansions for the mixed fields
\bea
\phi_A(x)&=&\int{d^3  k}\,\Big\{a_{\bk,A}(t)\,U_{\bk,1}(x)+\bar a^{\dagger}_{\bk,A}(t)\,U^*_{\bk,1}(x)\Big\},
\label{eqn:sub expansion bis a}\\[1mm]
\phi_B(x)&=&\int{d^3  k}\,\Big\{a_{\bk,B}(t)\,U_{\bk,2}(x)+\bar a^{\dagger}_{\bk,B}(t)\,U^*_{\bk,2}(x)\Big\},
\label{eqn:sub expansion bis b}
\eea
where the flavor operator $a_{\bk,A}$ is given by the KG product\hspace{0.3mm}\footnote {In Ref.\cite{bosonmix} the flavor operators are derived by means of the algebraic generator of the mixing relations Eqs. (\ref{Ponteca}),(\ref{Pontecb}). Here, we use this alternative, but completely equivalent approach.}:
\begin{equation}
a_{\bk,A}(t)=\left(\phi_A, U_{\bk,1}\right).
\label{eqn:aka}
\end{equation}
Similar expressions hold for the other flavor operators (see below). By inserting the expansion Eq. (\ref{Ponteca}) in the product Eq. (\ref{eqn:aka}) and exploiting the orthonormality condition of the plane-waves Eq. (\ref{eqn:minkowskinorm}), we obtain for $a_{\bk,A}$\footnote{To simplify the notation, in what follows the time dependence of flavor operators will be omitted when not necessary.}
\begin{equation}
a _{{\bf k},A}=\cos\theta\, a _{{\bf k},1}+\sin\theta\, \left({{\rho}^{\bk\, *}_{12}}\, a _{{\bf k},2}\, + \, {{\lambda}^\bk_{12}}\, \bar a^\dagger _{-{\bf k},2}\right),
\label{eqn:flavor-a}
\end{equation}
where the coefficients ${\tilde{\rho}^\bk_{12}}$ and ${\tilde{\lambda}^\bk_{12}}$ are given by
\begin{equation}
\label{eqn:bogolcoeffs}
{{\rho}^\bk_{12}}=|{{\rho}^\bk_{12}}|\, e^{i\left(\omega_{\bk,2}-\omega_{\bk,1}\right) t},\qquad {{\lambda}^\bk_{12}}=|{{\lambda}^\bk_{12}}|\,e^{i\left(\omega_{\bk,1}+\omega_{\bk,2}\right) t},\!\
\end{equation}
with
\begin{equation}
|{{\rho}^\bk_{12}}|\equiv\frac{1}{2}\left( \sqrt{\frac{\omega_{\bk,1}}{\omega_{\bk,2}}}+\sqrt{\frac{\omega_{\bk,2}}{\omega_{\bk,1}}}\right),\qquad |{{\lambda}^\bk_{12}}|\equiv\frac{1}{2}\left( \sqrt{\frac{\omega_{\bk,1}}{\omega_{\bk,2}}}-\sqrt{\frac{\omega_{\bk,2}}{\omega_{\bk,1}}}\right).
\label{eqn:bogomoduli}
\end{equation}
Similarly, for the other flavor operators we have
\bea
\label{eqn:barak}
\bar a _{{\bf k},A}&=&-\,{\big(\phi_A, U^*_{\bk,1}\big)}^\dagger=\cos\theta\, \bar a _{{\bf k},1}+\sin\theta\, \left({{\rho}^{\bk\, *}_{12}}\, \bar a _{{\bf k},2}\, + \, {{\lambda}^\bk_{12}}\,  a^\dagger _{{-\bf k},2}\right),
\\[2mm]
\label{eqn:ab}
a _{{\bf k},B}&=&{\left(\phi_B, U_{\bk,2}\right)}=\cos\theta\, a _{{\bf k},2}-\sin\theta\, \left(\bogub\, a _{{\bf k},1}\, - \, \bogvb\, \bar a^\dagger _{-{\bf k},1}\right),
\\[2mm]
\label{eqn:barab}
\bar a _{{\bf k},B}&=&-\,{\big(\phi_B, U^*_{\bk,2}\big)}^\dagger=\cos\theta\, \bar a _{{\bf k},2}-\sin\theta\, \left(\bogub\, \bar a _{{\bf k},1}\, - \, \bogvb\,  a^\dagger _{{-\bf k},1}\right).
\eea
Therefore, from Eqs. (\ref{eqn:flavor-a}), (\ref{eqn:barak})-(\ref{eqn:barab}), we realize that mixing relations within the quantum field theory framework arise as a consequence of the non-trivial interplay between   rotations and Bogolubov transformations at level of ladder operators. Such a structure, as known, has no corresponding result in quantum mechanics, where mixing transformations take the form of pure rotations operating on massive-particle states \cite{Pontecorvo}.

As it can be easily proved, the Bogolubov coefficients $|{{\rho}^\bk_{12}}|$ and $|{{\lambda}^\bk_{12}}|$ in Eq. (\ref{eqn:bogomoduli}) are related by
\begin{equation}
{|{{\rho}^\bk_{12}}|}^2-{|{{\lambda}^\bk_{12}}|}^2=1.
\label{eqn:u-vbos}\,
\end{equation}
The condition Eq. (\ref{eqn:u-vbos}) guarantees that flavor operators obey the canonical commutation relations (at equal times). Flavor vacuum at time $t$, denoted with $|0(\theta,t) \rangle_{A,B}$, is accordingly defined by
\begin{equation}
a _{{\bf k},\chi}(t)\;|0(\theta,t) \rangle_{A,B}=\bar a _{{\bf k},\chi}(t)\;|0(\theta,t) \rangle_{A,B}=0\hspace{0.2mm},\qquad \chi=A,B\hspace{0.2mm}.
\label{eqn:annichilasapori}
\end{equation}
The crucial point is that, in the infinite volume limit, flavor and mass vacua are found to be orthogonal to each other, thus giving rise to inequivalent Hilbert spaces\footnote{The unitary inequivalence between
representations of the canonical commutation relations is a characteristic feature of QFT, which is absent in quantum mechanics (QM) due to the von Neumann theorem \cite{Stone}.}. Indeed we have
\begin{equation}
\qquad\lim_{V \rightarrow \infty}\; _M\langle0|0(\theta,t)\rangle_{A,B}=0\hspace{0.2mm}, \qquad\forall t\hspace{0.2mm}.
\label{eqn:volumeinfinito}
\end{equation}
As shown in Ref.\cite{Gargiulo}, the responsible for such an inequivalence is the non commutativity between rotation and Bogoliubov transformation in Eq. (\ref{eqn:flavor-a}). The latter transformation, in particular, induces a drastic change into the vacuum structure, which becomes a condensate of particle/antiparticle pairs of density:
\begin{equation}
\label{eqn:denscondbosoni2}
{}_{A,B}\langle 0(\theta,t)| a_{{\bf k},i}^{\dagger}\, a_{{\bf
k},i} |0(\theta,t)\rangle_{A,B}= \sin^{2}\theta\; {|\bogvb|}^2, \qquad i=1,2\hspace{0.2mm}.
\end{equation}
A more detailed discussion about flavor vacuum structure can be found in Ref.\cite{bosonmix}.

The foregoing studies have been carried out by using the conventional plane-wave basis. By virtue of Lorentz covariance, however, field--mixing formalism can be equivalently analyzed within a basis diagonalizing any other Lorentz-group generator. For our purpose, for instance,  we may wonder  how the transformations Eqs. (\ref{eqn:flavor-a}), (\ref{eqn:barak})-(\ref{eqn:barab}) appear in the hyperbolic scheme, that is, the scheme which diagonalizes the boost operator (see Section \ref{Hyperbrepr}). To this end, by exploiting the completeness of the modes $\Big\{\widetilde{U}_{\kappa,i}^{\,(\sigma)},\,\widetilde{U}_{\kappa,i}^{\,(\sigma)*} \Big\}$, $i=1,2$ in Eq. (\ref{eqn:Uwidetilde}), let us adopt for mixed fields the hyperbolic free fields-like expansions Eq. (\ref{eqn:expansionfieldutilde})
\bea
\label{eqn:expancampmix}
\phi_{A}(x)&=& \sum_{\sigma,\hspace{0.4mm}\Omega}\,\int d^{2} {k}\,\Big\{d_{\kappa,A}^{\,(\sigma)}\;\widetilde{U}_{\kappa,1}^{\,(\sigma)}(x)+\bar d_{\kappa,A}^{\,(\sigma)\dagger}\;\widetilde{U}_{\kappa,1}^{\,(\sigma)*}(x)\Big\}\hspace{0.2mm},
\label{eqn:sub_expancampmix_a}\\
\nonumber\\
\phi_{B}(x)&=& \sum_{\sigma,\hspace{0.4mm}\Omega}\,\int d^{2} k\,\Big\{ d_{\kappa,B}^{\,(\sigma)}\;\widetilde{U}_{\kappa,2}^{\,(\sigma)}(x)+\bar d_{\kappa,B}^{\,(\sigma)\dagger}\;\widetilde{U}_{\kappa,2}^{\,(\sigma)*}(x)\Big\}\hspace{0.2mm},
\label{eqn:sub expancampmix b}
\eea
where the shorthand notation Eq. (\ref{eqn:simpnot}) has been used. Flavor operators in hyperbolic representation  are given by
\begin{eqnarray}
\label{eqn:daoperator}
\hspace{3mm}d_{\kappa,A}^{\,(\sigma)}&=&
\Big(\phi_A,\,\widetilde{U}_{\kappa,1}^{\,(\sigma)}\Big)
\,=\,\cos\theta\, d_{\kappa,1}^{\,(\sigma)}
\,+\,\sin\theta \sum_{\sigma',\hspace{0.4mm}\Omega'}\,\Big(d_{(\Omega',\vec{k}),2}^{\,(\sigma')}\;\bogocoeffAlpha\, +\,\bar d_{(\Omega',-\vec{k}),2}^{\,(\sigma')\dagger}\;\bogocoeffBeta\,\Big),\\
\nonumber\\[1mm]
\label{eqn:dabaroperator}
\bar d_{\kappa,A}^{\,(\sigma)}&=&
-\,{\Big(\phi_A,\,\widetilde{U}_{\kappa,1}^{\,(\sigma)*}\Big)}^\dagger
\,=\, \cos\theta\, \bar d_{\kappa,1}^{\,(\sigma)}
\,+\,\sin\theta \sum_{\sigma',\hspace{0.4mm}\Omega'}\,\Big(\bar d_{(\Omega',\vec{k}),2}^{\,(\sigma')}\;\bogocoeffAlpha\, +\,d_{(\Omega',-\vec{k}),2}^{\,(\sigma')\dagger}\;\bogocoeffBeta\,\Big),\\
\nonumber\\[1mm]
\label{eqn:dboperator}
\hspace{3mm}d_{\kappa,B}^{\,(\sigma)}&=&\Big(\phi_B,\,\widetilde{U}_{\kappa,2}^{\,(\sigma)}\Big)\,=\,\cos\theta\, d_{\kappa,2}^{\,(\sigma)}
-\,\sin\theta\sum_{\sigma',\hspace{0.4mm}\Omega'}\,\Big(d_{(\Omega',\vec{k}),1}^{\,(\sigma')}\;\bogocoeffAlphamensigstarkapduestarbi  -\,\,\bar d_{(\Omega',-\vec{k}),1}^{\,(\sigma')\,\dagger}\;\bogocoeffBetamensigstarkapduestarb\,\Big),\\
\nonumber\\[1mm]
\label{eqn:dbbaroperator}
\bar d_{\kappa,B}^{\!\ (\sigma)}&=&-\,{\Big(\phi_B,\,\widetilde{U}_{\kappa,2}^{\,(\sigma)\,*}\Big)}^\dagger=\cos\theta\, \bar d_{(\Omega,\vec{k}),2}^{\,(\sigma)}
-\,\sin\theta\sum_{\sigma',\hspace{0.4mm}\Omega'}\,\Big(\bar d_{(\Omega',\vec{k}),1}^{\,(\sigma')}\;\bogocoeffAlphamensigstarkapduestarbi \, -\, d_{(\Omega',-\vec{k}),1}^{\,(\sigma')\,\dagger}\;\bogocoeffBetamensigstarkapduestarb\,\Big).
\end{eqnarray}
As in the plane--wave framework, mixing relations exhibit the structure of a rotation combined with a  Bogolubov transformation. The Bogolubov coefficients $\bogocoeffAlpha$ and $\bogocoeffBeta$ are such that
\begin{equation}
\label{eqn:bogocoeffAB}
\Big(\widetilde{U}_{(\Omega',\vec{k}'),2}^{\,(\sigma')}\,,\,\widetilde{U}_{(\Omega,\vec{k}),1}^{\,(\sigma)}\Big)=\bogocoeffAlpha\,\,\delta^{2}(\vec{k}-\vec{k}'),\qquad\Big(\widetilde{U}_{(\Omega',\vec{k}'),2}^{\,(\sigma')*}\,,\,\widetilde{U}_{(\Omega,\vec{k}),1}^{\,(\sigma)}\Big)=\bogocoeffBeta\,\,\delta^{2}(\vec{k}+\vec{k}').
\end{equation}
They are given by the following expressions
\bea
&& \bogocoeffAlpha =\int^{+\infty}_{-\infty}\frac{dk_1}{4\pi}\lf(\frac{1}{\omega_{\bk,1}}+\frac{1}{\omega_{\bk,2}}\ri){\lf(\frac{\omega_{\bk,1}+k_1}{\omega_{\bk,1}-k_1}\ri)}^{i\sigma\Omega/2} {\lf(\frac{\omega_{\bk,2}+k_1}{\omega_{\bk,2}-k_1}\ri)}^{-i\sigma'\Omega'/2}
e^{i(\omega_{\bk,1}-\,\omega_{\bk,2})t},
\label{eqn:coefficientuno}
\\ [4mm]
&& \bogocoeffBeta =\int^{+\infty}_{-\infty}\frac{dk_1}{4\pi}\lf(\frac{1}{\omega_{\bk,2}}
-\frac{1}{\omega_{\bk,1}}\ri){\lf(\frac{\omega_{\bk,1}+
k_1}{\omega_{\bk,1}-k_1}\ri)}^{i\sigma\Omega/2} {\lf(\frac{\omega_{\bk,2}+k_1}{\omega_{\bk,2}-k_1}\ri)}^{-i\sigma'\Omega'/2}
e^{i(\omega_{\bk,1}+\,\omega_{\bk,2})t},
\label{eqn:coefficientdue}
\eea
which are to be compared with the corresponding relations in plane--wave representation Eqs. (\ref{eqn:bogolcoeffs}), (\ref{eqn:bogomoduli}). For $m_1\rightarrow m_2$, it is immediate to verify that $\bogocoeffBeta$ correctly vanishes, as it could be expected from Eqs. (\ref{eqn:ortonormbis}) and (\ref{eqn:bogocoeffAB}). Similarly, by exploiting the orthonormality property of $p$-functions Eq. (\ref{eqn:compl-p}), one can prove that $\bogocoeffAlpha\rightarrow \delta(\sigma-\sigma')\,\delta(\Omega-\Omega')$, according to Eqs. (\ref{eqn:ortonormbis}) and (\ref{eqn:bogocoeffAB}).

The analytical resolution of the integrals Eqs. (\ref{eqn:coefficientuno}), (\ref{eqn:coefficientdue}) is non--trivial\footnote{Note that the factorization of $\bogocoeffAlpha\Big|_{{}_{0}}$ and $\bogocoeffBeta\Big|_{{}_{0}}$ into the product of a Dirac delta function of $\sigma\Omega$ and $\sigma'\Omega'$ with suitable coefficients adopted in Ref.\cite{Proceeding}, turns out to be not properly correct.}. In spite of these technical difficulties, however, a reliable approximation can be obtained for $t=0$; in this case, indeed, Eqs. (\ref{eqn:coefficientuno}), (\ref{eqn:coefficientdue}) reduce to
\bea
&& \bogocoeffAlpha\Big|_{{}_{0}} =\int^{+\infty}_{-\infty}\frac{dk_1}{4\pi}\lf(\frac{1}{\omega_{\bk,1}}+\frac{1}{\omega_{\bk,2}}\ri){\lf(\frac{\omega_{\bk,1}+k_1}{\omega_{\bk,1}-k_1}\ri)}^{i\sigma\Omega/2} {\lf(\frac{\omega_{\bk,2}+k_1}{\omega_{\bk,2}-k_1}\ri)}^{-i\sigma'\Omega'/2}\,,
\label{eqn:coefficientunobistzeroappendix}
\\ [4mm]
&& \bogocoeffBeta\Big|_{{}_{0}} =\int^{+\infty}_{-\infty}\frac{dk_1}{4\pi}\lf(\frac{1}{\omega_{\bk,2}}
-\frac{1}{\omega_{\bk,1}}\ri){\lf(\frac{\omega_{\bk,1}+
k_1}{\omega_{\bk,1}-k_1}\ri)}^{i\sigma\Omega/2} {\lf(\frac{\omega_{\bk,2}+k_1}{\omega_{\bk,2}-k_1}\ri)}^{-i\sigma'\Omega'/2}\,.
\label{eqn:coefficientduebistzeroappendix}
\eea
In the realistic limit of small mass difference $\frac{\Delta m^2}{m_i^2}\equiv \frac{m_2^2-m_1^2}{m_i^2}\ll 1$, $i=1,2$, by using the expansions:
\begin{eqnarray}
\label{eqn:approx1}
\frac{1}{\omega_{\bk,2}}&=&\frac{1}{\omega_{\bk,1}}\,-\,\frac{1}{2}\,\frac{\Delta m^2}{\omega_{\bk,1}^{\hspace{0.3mm}3}}\,+\,\frac{3}{8}\,\frac{{(\Delta m^2)}^2}{\omega_{\bk,1}^{\hspace{0.3mm}5}}\,+\,
\mathcal{O}\left({\frac{{(\Delta m^2)}^3}{\omega_{\bk,1}^{\hspace{0.3mm}7}}}\right),\\[3mm]
\label{eqn:approx2}
\frac{\omega_{\bk,2}}{\omega_{\bk,1}^{\hspace{0.3mm}2}}&=&\frac{1}{\omega_{\bk,1}}\,+\,\frac{1}{2}\,\frac{\Delta m^2}{\omega_{\bk,1}^{\hspace{0.3mm}3}}\,-\,\frac{1}{8}\,\frac{{(\Delta m^2)}^2}{\omega_{\bk,1}^{\hspace{0.3mm}5}}\,+\,\mathcal{O}\left({\frac{{(\Delta m^2)}^3}{\omega_{\bk,1}^{\hspace{0.3mm}7}}}\right),
\end{eqnarray}
and the formula:
\begin{equation}
\label{eqn:formula}
\int_{-\infty}^{+\infty}\frac{dy}{\cosh^2y}\, \cos\left[\sigma\,(\Omega'-\Omega)\,y\right]=\frac{\pi\,\sigma\,(\Omega'-\Omega)}{\sinh\left[\frac{\pi}{2}\,\sigma\,(\Omega'-\Omega)\right]},
\end{equation}
the following leading--order approximations can be derived for $\bogocoeffAlpha\Big|_{{}_{0}}$ and $\bogocoeffBeta\Big|_{{}_{0}}$
\bea
\bogocoeffAlpha\Big|_{{}_{0}}&=&\delta_{\sigma\sigma'}\,\delta{(\kappa-\kappa')}-\frac{\Delta m^2}{8\mu_{k,1}^{{\hspace{0.3mm}}2}}\,\frac{\sigma\,\Omega-\sigma'\Omega'}{\sinh[\frac{\pi}{2}(\sigma\,\Omega-\sigma'\Omega')]}\,+\,\mathcal{O}\left({{\Big(\frac{\Delta m^2}{{\mu_{k,1}^{\hspace{0.3mm}2}}}\Big)}^2}\right),
\label{eqn:coefficientunoapprox}
\\ [4mm]
\bogocoeffBeta\Big|_{{}_{0}}&=&-\frac{\Delta m^2}{8\mu_{k,1}^{{\hspace{0.2mm}}2}}\,\frac{\sigma\,\Omega-\sigma'\Omega'}{\sinh[\frac{\pi}{2}(\sigma\,\Omega-\sigma'\Omega')]}\,+\,\mathcal{O}\left({{\Big(\frac{\Delta m^2}{{\mu_{k,1}^{\hspace{0.2mm}2}}}\Big)}^2}\right).
\label{eqn:coefficientdueapprox}
\eea

To conclude, let us remark that Eqs. (\ref{eqn:daoperator})-(\ref{eqn:dbbaroperator}) are canonical transformations (at equal times): indeed, by exploiting the commutators of $d_{\kappa,i}^{\,(\sigma)}$ and $\bar d_{\kappa,i}^{\,(\sigma)}$ ($i=1,2$) in Eq. (\ref{eqn:commut-d}) and the relations Eqs. (\ref{eqn:coefficientuno}), (\ref{eqn:coefficientdue}), it can be verified that
\begin{equation}
\label{eqn:commutatorflavor}
\left[d_{\kappa,\chi}^{\,(\sigma)}(t)\,,\, d_{\kappa',\chi'}^{ ( \sigma')\dagger}\,(t')\right]\bigg|_{t=t'}=\left[\bar d_{\kappa,A}^{\,(\sigma)}\,(t)\,,\, {\bar d_{\kappa',A}}^{\,(\sigma')\dagger}\,(t')\right]\bigg|_{t=t'}\,=\,\delta_{\chi\chi'}\,\delta_{\sigma\sigma'}\,\delta^3(\kappa-\kappa'),\,\qquad \chi,\chi'=A,B,
\end{equation}
with all other commutators vanishing. The crucial point is that Eq. (\ref{eqn:commutatorflavor}) holds -- and can be explicitly proved -- in spite of the technical difficulties in the evaluation of $\bogocoeffAlpha$ and $\bogocoeffBeta$. Here, for clarity, we report the calculation of $\left[d_{\kappa,A}^{\,(\sigma)}\,(t)\,,\, d_{\kappa',A}^{ ( \sigma')\dagger}\,(t')\right]$ for $t=t'=0$ (similarly for $\left[d_{\kappa,B}^{\,(\sigma)}\,(t)\,,\, d_{\kappa',B}^{ ( \sigma')\dagger}\,(t')\right]$)
\begin{eqnarray}
\nonumber
&&\hspace{-6mm}\left[d_{\kappa,A}^{\!\ (\sigma)}\,,\,d_{\kappa',A}^{\!\ (\sigma')\,\dagger}\right]\bigg|_{0}=\\[4mm]
\nonumber
&&\hspace{-6mm}\Bigg[\cos\theta\, d_{\kappa,1}^{\!\ (\sigma)}\hspace{0.2mm}+\hspace{0.2mm}\sin\theta\hspace{-0.8mm}\sum_{\hspace{0.6mm}\sigma'',\,\Omega''}\Bigg\{d_{(\Omega'',\vec{k}),2}^{\!\ (\sigma'')}\,\int\hspace{-0.7mm}\frac{dk_1}{4\pi}\;g_+(k_1)\;\frac{{f^{\,(\sigma)}\,(\omega_{\textbf{k},1},\Omega)}}{{f^{\,(\sigma'')}\,(\omega_{\textbf{k},2},\Omega^{''})}}\hspace{0.2mm}+\hspace{0.2mm}\bar d_{(\Omega'',-\vec{k}),2}^{\!\ (\sigma'')\,\dagger}\,\int\hspace{-0.7mm}\frac{dk_1}{4\pi}\;g_{-}(k_1)\;\frac{{f^{\,(\sigma)}\,(\omega_{\textbf{k},1},\Omega)}}{{f^{\,(\sigma'')}\,(\omega_{\textbf{k},2},\Omega'')}}\Bigg\}\,,\\[4mm]
\nonumber
&&\hspace{-5mm}\cos\theta\, d_{\kappa',1}^{\!\ (\sigma')\,\dagger}\hspace{0.2mm}+\hspace{0.2mm}\sin\theta\hspace{-1.7mm}\sum_{\hspace{0.6mm}\sigma''',\,\Omega'''}\hspace{-1mm}\Bigg\{d_{(\Omega''',\vec{k}'),2}^{\!\ (\sigma''')\,\dagger}\,\int\hspace{-0.7mm}\frac{dk'_1}{4\pi}\;g_+(k'_1)\;\frac{{f^{\,(\sigma''')}\,(\omega_{\textbf{k}',2},\Omega^{'''})}}{{f^{\,(\sigma')}\,(\omega_{\textbf{k}',1},\Omega^{'})}}\hspace{0.2mm}+\hspace{0.2mm}\bar d_{(\Omega''',-\vec{k}'),2}^{\!\ (\sigma''')}\,\int\hspace{-0.7mm}\frac{dk'_1}{4\pi}\;g_{-}(k'_1)\,\frac{{f^{\,(\sigma''')}\,(\omega_{\textbf{k}',2},\Omega^{'''})}}{{f^{\,(\sigma')}\,(\omega_{\textbf{k}',1},\Omega^{'})}}\Bigg\}\Bigg],
\end{eqnarray}
where we introduced the shorthand notation in Eq. (\ref{eqn:simpnot}) and adopted the following definitions
\begin{equation}
\hspace{10mm}g_{\pm}(k_1)=\frac{1}{\omega_{\textbf{k},2}}\pm\frac{1}{\omega_{\textbf{k},1}},\qquad f^{\,(\sigma)}\,(\omega_{\textbf{k},j},\Omega)={\left(\frac{\omega_{\textbf{k},j}+k_1}{\omega_{\textbf{k},j}-k_1}\right)}^{i\,\sigma\,\Omega/2}, \qquad j=1,2.
\end{equation}
By exploiting the commutation relations Eq. (\ref{eqn:commut-d}), it follows that
\begin{eqnarray}
\nonumber
\hspace{-1mm}\left[d_{\kappa,A}^{\!\ (\sigma)}\,,\,\,d_{\kappa',A}^{\!\ (\sigma')\dagger}\right]\bigg|_{0}&=&\cos^2\theta\,\delta_{\sigma\sigma'}\,\delta^3(\kappa-\kappa')\\[3mm]
\nonumber
&+&\,\sin^2\theta\hspace{-0.6mm}\sum_{\hspace{0.6mm}\sigma'',\hspace{0.1mm}\Omega''}\Bigg\{\int\frac{dk_1}{4\pi}\;g_{+}(k_1)\;\frac{{f^{\,(\sigma)}\,(\omega_{\textbf{k},1},\Omega)}}{{f^{\,(\sigma'')}\,(\omega_{\textbf{k},2},\Omega^{''})}}\,\int\frac{dk'_1}{4\pi}\;g_{+}(k'_1)\;\frac{{f^{\,(\sigma'')}\,(\omega_{\textbf{k}',2},\Omega^{''})}}{{f^{\,(\sigma')}\,(\omega_{\textbf{k}',1},\Omega^{'})}}\Bigg\}\,\delta\hspace{0.2mm}^{2}\,(\vec{k}-\vec{k}')\\[4mm]
\nonumber
&-&\sin^2\theta\hspace{-0.6mm}\sum_{\hspace{0.6mm}\sigma'',\hspace{0.1mm}\Omega''}\Bigg\{\int\frac{dk_1}{4\pi}\;g_{-}(k_1)\;\frac{{f^{\,(\sigma)}\,(\omega_{\textbf{k},1},\Omega)}}{{f^{\,(\sigma'')}\,(\omega_{\textbf{k},2},\Omega'')}}\,\int\frac{dk'_1}{4\pi}\;g_{-}(k'_1)\;\frac{{f^{\,(\sigma'')}\,(\omega_{\textbf{k}',2},\Omega^{''})}}{{f^{\,(\sigma')}\,(\omega_{\textbf{k}',1},\Omega^{'})}}\Bigg\}\,\delta\hspace{0.2mm}^{2}\,(\vec{k}-\vec{k}').
\end{eqnarray}
In definitive, by means of Eqs. ({\ref{eqn:completeorthonorm-p}}) and (\ref{eqn:compl-p}), we have the result:
\begin{eqnarray}
\nonumber
\left[d_{\kappa,A}^{\!\ (\sigma)}\,,\,d_{\kappa',A}^{\!\ (\sigma')\dagger}\right]\bigg|_{0}&=&\cos^2\theta\,\delta_{\sigma\sigma'}\,\delta^3(\kappa-\kappa')\\[3mm]
\nonumber
&+&\frac{1}{2}\,\sin^2\theta\int\frac{dk_1}{2\pi\omega_{\textbf{k},1}}\Bigg\{\int dk'_1\,\bigg[\,\frac{{f^{\,(\sigma)}\,(\omega_{\textbf{k},1},\Omega)}}{{f^{\,(\sigma')}\,(\omega_{\textbf{k}',1},\Omega^{'})}}\,\underbrace{\hspace{-0.6mm}\sum_{\hspace{0.6mm}\sigma'',\hspace{0.1mm}\Omega''}\frac{1}{2\pi\omega_{\textbf{k}',2}}\,\frac{{f^{\,(\sigma'')}\,(\omega_{\textbf{k}',2},\Omega^{''})}}{{f^{\,(\sigma'')}\,(\omega_{\textbf{k},2},\Omega^{''})}}}_{\delta(k_1-k_1')}\bigg]\Bigg\}\;\delta\hspace{0.2mm}^{2}({\vec{k}-\vec{k}'})\\[4mm]
\nonumber
&+&\frac{1}{2}\,\sin^2\theta\int dk_1\,\Bigg\{\int\frac{dk'_1}{2\pi\omega_{\textbf{k}',1}}\,\bigg[\,\frac{{f^{\,(\sigma)}\,(\omega_{\textbf{k},1},\Omega)}}{{f^{\,(\sigma')}\,(\omega_{\textbf{k}',1},\Omega')}}\,\underbrace{\hspace{-0.6mm}\sum_{\hspace{0.6mm}\sigma'',\hspace{0.1mm}\Omega''}\frac{1}{2\pi\omega_{\textbf{k},2}}\;\frac{{f^{\,(\sigma'')}\,(\omega_{\textbf{k}',2},\Omega^{''})}}{{f^{\,(\sigma'')}\,(\omega_{\textbf{k},2},\Omega^{''})}}}_{\delta(k_1-k_1')}\bigg]\Bigg\}\,\delta\hspace{0.2mm}^{2}({\vec{k}-\vec{k}'})\\[4mm]
\nonumber
&=&\cos^2\theta\,\delta_{\sigma\sigma'}\,\delta^3(\kappa-\kappa')\,+\,\sin^2\theta\underbrace{\int\frac{dk_1}{2\pi\omega_{\textbf{k},1}}\;\frac{{f^{\,(\sigma)}\,(\omega_{\textbf{k},1},\Omega)}}{{f^{\,(\sigma')}\,(\omega_{\textbf{k}',1},\Omega')}}}_{\delta_{\sigma\sigma'}\,\delta(\Omega-\Omega')}\,\delta\hspace{0.2mm}^{2}(\vec{k}-\vec{k}')\,=\,\delta_{\sigma\sigma'}\,\delta^{3}(\kappa-\kappa').
\end{eqnarray}

The canonical commutators Eq. (\ref{eqn:commutatorflavor}) allow us to derive the following useful conditions for the Bogolubov coefficients $\bogocoeffAlpha\Big|_{{}_{0}}$ and $\bogocoeffBeta\Big|_{{}_{0}}$
\begin{eqnarray}
\label{eqn:primprop}
\sum_{\sigma'',\hspace{0.4mm}\Omega''}\left({\cal A}_{(\Om',\Om''),\,\vec{k}'}^{(\si',-\si'')}\;\,{\cal B}_{(\Om,\Om''),\,\vec{k}}^{(\si,\si'')\,*}\,-\,{\cal A}_{(\Om,\Om''),\,\vec{k}}^{(\si,\si'')\,*}\;\,{\cal B}_{(\Om',\Om''),\,\vec{k}'}^{(\si',-\si'')}\right)\Big|_{{}_{0}}&=&0,\hspace{9mm}\\[3mm]
\label{eqn:secprop}
\sum_{\sigma'',\hspace{0.4mm}\Omega''}\left(\bogocoeffAlphater\;\,\bogocoeffAlphaquintst\,-\,\bogocoeffBetater\;\,\bogocoeffBetaquintst\right)\Big|_{{}_{0}}&=&\delta_{\sigma\sigma'}\delta^3(\kappa-\kappa').
\end{eqnarray}
The latter one, in particular, is the hyperbolic representation equivalent of Eq. (\ref{eqn:u-vbos}).

\end{document}